\documentclass{article}

\usepackage{arxiv}

\usepackage[utf8]{inputenc} 
\usepackage[T1]{fontenc}    
\usepackage{hyperref}       
\usepackage{url}            
\usepackage{amsmath}
\usepackage{booktabs}       
\usepackage{amsfonts}       
\usepackage{nicefrac}       
\usepackage{microtype}      
\usepackage{lipsum}
\usepackage{authblk}
\usepackage{xcolor}
\usepackage{graphicx}
\usepackage{float}
\usepackage{tabularx}
\usepackage{tcolorbox}
\usepackage{subcaption}

\title{\Large \bf {LED there be DoS}: Exploiting variable bitrate {IP} cameras for network {DoS}}

\author[1,2]{\small Emmanuel Goldberg}
\author[1,2]{\small Oleg Brodt}
\author[2,3]{\small Aviad Elyashar}
\author[1,2]{\small Rami Puzis}
\affil[1]{Department of Software and Information Systems Engineering, Ben-Gurion University of the Negev, Israel}
\affil[2]{Cyber Labs@BGU,Ben-Gurion University of the Negev, Israel}
\affil[3]{Department of Computer Science, Shamoon College of Engineering, Israel}

\begin{document}
\maketitle

\begin{abstract}
Variable-bitrate video streaming is ubiquitous in video surveillance and CCTV, enabling high-quality video streaming while conserving network bandwidth. 
However, as the name suggests, variable-bitrate IP cameras can generate sharp traffic spikes depending on the dynamics of the visual input.
In this paper, we show that the effectiveness of video compression can be reduced by up to $6 \times$ using a simple laser LED pointing at a variable-bitrate IP camera, forcing the camera to generate excessive network traffic.
Experiments with IP cameras connected to wired and wireless networks indicate that a laser attack on a single camera can cause significant packet loss in systems sharing the network with the camera and reduce the available bandwidth of a shared network link by 90\%. 
This attack represents a new class of cyber-physical attacks that manipulate variable bitrate devices through changes in the physical environment without a digital presence on the device or on the network.
We also analyze the broader view of multidimensional cyberattacks that involve both the physical and digital realms and present a taxonomy that categorizes attacks based on their direction of influence (physical-to-digital or digital-to-physical) and their method of operation (environment-driven or device-driven), highlighting multiple areas for future research.
  
\end{abstract}


\section{Introduction}
\label{sec:introduction}

Video streaming cameras are an integral part of modern life. 
Scattered across cities~\cite{hino2022changes}, office buildings~\cite{rondon2022survey}, transportation hubs~\cite{leone2021survey}, and critical facilities~\cite{gyorgyi2023network}, they monitor the environment for hazards, suspicious activities, quality control, etc.
In the past, such cameras relied on dedicated CCTV infrastructure, using specialized equipment and coaxial cabling. 
However, modern surveillance cameras are network-connected, embracing IP-based technologies.
While the adoption of IP-based technologies has increased efficiency and reduced costs by allowing cameras to connect to existing IP networks, it has also introduced security problems inherent to IP networks \cite{kalbo2020security}.

IP cameras first compress the raw video using codecs like H.264, AV1, or H.265 to reduce data size.
The compressed video payload is then encapsulated with protocol headers, such as RTP (for streaming), and transport-layer headers (typically UDP), followed by IP headers~\cite{mohamed2023survey}.
Transmitting full high-resolution frames at high frequencies (e.g., 30 or 60 frames per second (FPS)) for a static scene is wasteful.
In addition to compressing individual frames, video codecs in IP cameras reduce bandwidth usage by transmitting mainly the changes in the captured scene~\cite{li2021modern}.
Modern surveillance cameras operate under the assumption that the scene they capture remains mostly static over time. 
A high-definition (HD) IP camera would typically produce 1.5-3.0 Mbps for effective surveillance and object detection~\cite{xu2015much,gravzelisminimum}. 
In a typical office environment, we observed around 2-2.5 Mbps of traffic generated by a full HD camera with a 60 FPS stream using H.265 encoding.
The assumption of scene stationarity fundamentally shapes network architecture and bandwidth planning. 
An attack we investigate in this paper challenges this assumption. 
By introducing rapid changes to the scene viewed by the camera, an attacker can force the camera to generate excessive network traffic.
We explore a setup where an attacker uses a flickering laser diode to generate rapid changes in the scene.  
We evaluate the attack in laboratory settings and show
that it can increase the bandwidth consumption of a camera by a factor of $5-6\times$ 
potentially overloading the communication network. 
We show a significant increase in packet drop rate and round trip time in wired and wireless network setups as a result of exploiting a single IP camera. 

Our contributions are as follows:
\begin{itemize} 
\item We demonstrate a new category of attacks where an IoT device (IP camera) is used to attack its network without digital interaction with the device. 
\item We demonstrate how the adaptive bitrate of an IP camera is exploited to generate excessive network traffic and evaluate the potential impacts of such an attack.
\item We survey the physical-digital and digital-physical spillovers in the context of cybersecurity, present a taxonomy and identify multiple gaps in physical-to-digital attacks.
\end{itemize}

\section{Background and Related Work}
\label{sec:background}

Simply put, IP cameras are network-connected devices that convert physical movement in their field of view into packets. 
The more movement in the scene, the more packets the cameras generate and transmit to the network. In this section, we provide background material on generating a network denial-of-service (DoS) attack with IP cameras, as well as relevant related work.

\subsection{IP Cameras as Network Devices}
\label{sec:compression}
Modern digital cameras convert light into electronic signals to capture images or video footage. Each video recording is a sequence of frames, typically captured at rates of 24-120 FPS, which, when displayed in sequence to the human eye, create the illusion of motion. These frames generate substantial raw data. For example, an uncompressed 1080×1920 video at 60 FPS requires $\approx 2.986Gb/s$, making raw storage and transmission impractical. To address this, video compression techniques reduce the data size while maintaining a balance between visual quality and computational efficiency \cite{Mitrovic_VideoCompression, jack2011video, nassi2018gamedronesdetecting}.

Video encoding takes advantage of the fact that our eyes are more sensitive to variations in brightness (luminance) than to changes in color (chrominance) in order to reduce the resolution of color information through a process called \textit{chroma subsampling}, where color data are sampled at a lower resolution than brightness data. Common subsampling schemes like 4:2:2 and 4:2:0 keep all of the luminance data while sampling chrominance at $1/2$ and $1/4$ resolution, respectively.

Another aspect of video encoding is exploiting redundancies within and between frames. Spatial redundancy arises from similarities between neighboring pixels in a single frame, while temporal redundancy stems from similarities between consecutive frames. Techniques such as discrete cosine transform (DCT) address spatial redundancy by transforming image blocks into frequency components. High-frequency components, which represent fine details less noticeable to human viewers, are quantized or discarded. This reduces the data required to represent a frame.

Temporal redundancy arises because consecutive frames in a video sequence often contain similar content, especially in scenes with minimal motion. Instead of encoding each frame in full, only the changes between frames are encoded. This is achieved through motion estimation and compensation.

Regular motion events in an office environment have a negligible impact on the stream size, as modern video compression algorithms are based on motion vectors. 
Motion estimation identifies blocks of pixels in one frame corresponding to blocks in a reference frame. The displacement of these blocks is represented by \textit{motion vectors}, which describe the direction and magnitude of movement. Motion compensation uses these vectors to predict the content of the current frame based on reference frames. By encoding only the differences between the actual and predicted frames, the data size is significantly reduced.
This means that a person walking in a corridor has a similar impact on the generated traffic as a person standing in place. 
Therefore, to generate additional traffic, unexpected or unusual changes in the scene are required.


Video encoding employs three types of frames, which work together to balance compression efficiency and video quality:

\begin{itemize} 
    \item \textbf{I-Frames (Intra-coded):} These frames serve as independent reference points, and they are fully encoded without relying on other frames.
    \item \textbf{P-Frames (Predicted):} These frames only store the differences from a preceding frame, reducing data size.
    \item \textbf{B-Frames (Bidirectional):} These frames use data from preceding and subsequent frames to increase encoding efficiency.
\end{itemize}

When encoding a live video stream, B-frames are typically avoided, as the successive frames necessary to define a B-frame are in the yet-unknown future.

Modern video encoding algorithms typically perform a multi-stage process consisting of the following steps:
\begin{enumerate} 
\item \textbf{Color Space Conversion and Chroma Subsampling:} Convert the raw RGB data into a suitable format (e.g., YCbCr) and reduce chrominance resolution.
\item \textbf{Partitioning:} Divide each frame into blocks (e.g., $8\times8$ or $16\times16$ pixels) for efficient and independent processing.
\item \textbf{Motion Estimation and Compensation:} Identify motion vectors and encode only the changes between frames.
\item \textbf{Transformation and Quantization:} Apply DCT to reduce spatial redundancy, followed by quantization to lower precision when less detail is perceived.
\item \textbf{Entropy Coding:} Use lossless compression techniques such as Huffman coding and run-length encoding (RLE) to efficiently represent the quantized data.
\end{enumerate}

Therefore, for the attack to succeed, we need to render these optimizations as ineffective as we can. This is done by minimizing the temporal redundancies within each frame and the spatial redundancies between frames, increasing the data needed to encode key and delta frames.

\subsection{DDoS Attacks and IP Cameras}

As IoT continues to expand, IP cameras have increasingly become targets for attackers. Threat actors exploit IP cameras primarily through two categories of attacks, which we refer to as invasive and non-invasive attacks.

In invasive attacks, attackers gain unauthorized access to the internals of the device to obtain privileged capabilities, such as the capability to execute arbitrary code, modify the device’s configuration settings, or even replace firmware. These attacks are often facilitated by firmware vulnerabilities that allow remote code execution (RCE) \cite{alrawi2019ip}. A notable example is the Mirai botnet, which compromised IoT devices, including IP cameras, by brute-forcing weak credentials and implanting malware that executed attacker-controlled instructions \cite{antonakakis2017mirai}. 
Similarly, vulnerabilities in devices have been exploited to perform unauthenticated RCE, turning IP cameras into potential surveillance or attack tools \cite{watchfulip2021hikvision}. Additional studies have demonstrated how specific IoT camera models were susceptible to invasive exploits that allowed complete device control \cite{stabili2024tenda}.

In contrast, in non-invasive attacks, attackers exploit network-connected cameras without directly modifying the device's firmware or internal configurations. Instead, they leverage the device's existing functionalities to achieve unintended outcomes, such as participating in DDoS amplification attacks. This method does not require installing malware on the device. For instance, misconfigured IoT devices, including IP cameras, have been exploited in Simple Service Discovery Protocol (UPnP\textbackslash SSDP) amplification attacks, where attackers send spoofed requests to these devices, causing them to send large responses to the target \cite{kolias2017mirai, helpnetsecurity2019ssdp}.

In recent years, we have seen a rise in network defense technologies that specifically address DDoS attacks. 
To overload a network remotely, an attacker typically needs to generate a significant amount of traffic and bypass anti-DDoS scrubbing centers, which filter out DDoS traffic and pass only legitimate packets to the client. In 2024, Cloudflare, which provides an anti-DDoS cloud-based scrubbing service, mitigated the largest DDoS attack on record, peaking at 5.6 terabits per second (Tbps) and 666 million packets per second (pps) \cite{cloudflare_famous_ddos}. Since many critical processes are heavily filtered or even disconnected from the Internet, mounting a successful remote DDoS attack that impacts a critical process is practically infeasible.
Instead of targeting the network remotely, an attacker could connect to a LAN and launch a DoS attack internally. 
However, this is even more challenging, as the attacker would need to bypass physical security and network access controls.

\textbf{Unlike existing DoS attacks, our method requires neither remote nor local network access -- a line of sight to the surveillance cameras of the targeted facility is all that is needed.}

\subsection{Variable Bitrate Exploitation}

The phenomenon in which physical-world events directly influence network traffic patterns is prevalent among IoT devices, industrial control system (ICS) actuators and sensors, industrial IoT (IIoT) devices, and cyber-physical systems in general~\cite{xu2018survey}.
The ability to connect between the physical real world and the digital man-made realms is the raison d'être of cyber-physical system development. 
It is the ability to represent physical phenomena in digital terms and vice versa that gave rise to cyber-physical systems in the first place. Once you can translate the physical into the digital (e.g., by using sensors), you can also apply computation and automated decision-making. Such decisions can be translated back from the digital realm to physical actions (e.g., by using actuators), thereby interconnecting the physical and digital worlds.

In this context, the relationship between physical events and network traffic represents a fundamental design pattern in cyber-physical architectures, where computer systems typically monitor some real-world aspects and report on their observations via a network~\cite{mois2016cyber}. 
In IP networks, cyber-physical devices are sometimes designed to adapt their communication patterns based on real-world stimuli. This adaptation mechanism is important for efficient network utilization and system operation, as well as energy saving, which is especially critical for battery-powered devices. The philosophy behind adaptive behavior (e.g., adaptive data sampling) is that the behavior should depend on the rate at which the real-world phenomenon changes (e.g., signal changes) \cite{waterSampling2017}. 
Similarly, video and audio devices adjust their bitrate based on changes in the video or audio they capture. 
In this context, two techniques have emerged over the last decade to efficiently manage network bandwidth: adaptive sampling and adaptive filtering. 
These techniques dynamically adjust data transmission rates and filter thresholds based on environmental conditions \cite{giouroukis2020survey}. 
For example, when sensors detect significant changes or events of interest, they may increase their sampling and transmission rates~\cite{anta2010sample,molhem2024novel}, leading to higher volumes of network traffic. Conversely, during periods of stability or inactivity, these algorithms reduce data transmission to conserve network resources. 

In this regard, consider industrial vibration sensors that increase their sampling and transmission rates upon detecting anomalous machinery behavior~\cite{yin2023adaptive}. 
Under normal operating conditions, these sensors may transmit minimal data. However, when they detect unusual vibrations, they can flood the network with high-frequency measurements and alerts.
Some audio monitoring systems in industrial environments also demonstrate this characteristic, generating variable bitrate traffic that correlates with changes in noise levels \cite{AudioSampling2010}.

In cases where the sensors are battery-powered, which is quite common for in-field deployed sensors, increased activity may lead to quicker battery exhaustion \cite{Harb2018Energy-Efficient}. 
For example, to conserve sensor battery life, it is beneficial to reduce the amount of data sampled by water sensors when the monitored environment is stable. 
Submerged sensors that monitor water quality may increase their water sampling rate rapidly once a change in pH value, dissolved oxygen (DO), conductivity, oxidation-reduction potential (ORP), turbidity, or temperature is detected \cite{waterSampling2017}. When certain parameters of the water change abruptly, the sampling frequency automatically increases to gather enough information about these changes.

\textbf{While adaptive behavior optimizes power and bandwidth usage under normal operating conditions, it also results in the direct influence of physical world events on network utilization. 
This relationship creates a potentially exploitable connection between environmental stimuli and resource exhaustion, such as battery depletion.} 
Moreover, an adversary capable of manipulating the physical environment could exploit these adaptive behaviors, resulting in excessive network traffic, which may lead to decreased availability of other resources and services that rely on the network.

\section{Generating a DoS with a Laser}

We demonstrate how variable bitrate exploitation can be weaponized in the context of IP cameras, where deliberate changes in the visual scene can trigger a network DoS through the video encoding mechanism. 
This attack, which creates a DoS condition by manipulating the physical environment within the cameras' fields of view, not only illustrates the exploitation of variable bitrate but also highlights a broader security challenge in cyber-physical systems--one where network behavior can be maliciously influenced through controlled real-world stimuli.

\subsection{Threat Model}
In this attack, a threat actor would like to disturb communication in the victim network, causing a denial of service (DoS). 

\subsubsection{Attackers' Motivation}
Network communication disruptions, even if brief, can have serious consequences in ICS environments. Consider a scenario where an industrial process deviates from its operational thresholds--whether accidentally or not--and requires calibration. The process engineer may need to intervene manually through the human-machine interface (HMI) to adjust process parameters. However, if network dysfunction prevents the HMI from reliably transmitting control commands to the programmable logic computer (PLC) and eventually the field devices while the process remains outside normal parameters, the safety system may initiate an automatic shutdown sequence.
In this scenario, the network failure creates a loss-of-control situation where the safety instrumented system (SIS) forces the process to halt to prevent potential damage or hazardous conditions \cite{mitre_t0827}.
The financial impact of such disruptions can be non-negligible. In continuous process industries like oil refineries, chemical plants, or semiconductor manufacturing, an unplanned shutdown can result in significant production losses. In industries producing perishable goods, such as food and beverage processing, an unexpected halt can lead to spoilage of entire batches of in-process materials that are sensitive to time and temperature conditions. Besides direct losses, recovery from a safety shutdown is not immediate; it often requires systematic equipment checks, recalibration of instruments, and process restart procedures that can take hours or even days. Equipment stress from unplanned shutdowns can also lead to increased maintenance costs.

However, loss of control is not the only potential consequence of network disruptions. When a historian server becomes unreachable over the network, the facility loses its ability to log process data \cite{mitre_t0815}. This is more than merely a nuisance. In fact, in pharmaceutical manufacturing, where detailed batch records are required for regulatory compliance \cite{FDA_21CFR}, losing historian functionality could limit the ability to prove that products were manufactured under the required conditions. Since the regulations mandate that electronic records must be accurate, reliable, and readily retrievable \cite{FDA_21CFR}, the inability to fully document production parameters could result in entire batches being rejected (even if the manufacturing process itself remained within specifications) simply due to the lack of verifiable records. Similar regulations apply in other regulated sectors, such as food and chemicals, stressing that a single network disruption can have far-reaching financial and operational consequences \cite{ICHQ7}.

\subsubsection{Assumptions}
Following are the assumptions made about the attacker and the victim.
We assume that the attacker does not have and cannot acquire a digital foothold within the victim's network.  
But there is a clear line of sight, allowing the attacker to point a laser at one or more surveillance cameras operated by the victim at their premises. 
The cameras should remain at line of sight from an effective angle for the desired duration of the attack. 
The victim camera(s) is a typical IP camera using a standard variable-bitrate codec and connected to an IP network.
    
We assume that the victim operates a standard network with Ethernet switches operating (e.g., 100 Mb/s) and typical wireless access points (e.g., 450 Mb/s). 
The victim network may have heterogeneous devices operating at different communication and processing speeds. 
The wired bandwidth is not over-provisioned by an order of magnitude and may reach 80\%-90\% capacity at peak times. 

We assume there is logical segmentation of the network such that the cameras can communicate only with certain dedicated hosts but not with any other devices in the network.
The video stream is transmitted through the same network devices (switches, routers, or access points) as the critical traffic -- no physical segmentation. 
Finally, we assume that the victim did not deploy traffic shaping or quality of service controls that limit the traffic emanating from cameras before it reaches the network bottlenecks.

\subsubsection{Attack Flow}
Given these assumptions, 
the adversary may execute the attack by pointing a laser flicker beam toward one or more IP cameras within the victim's premises.
The affected cameras identify changes in lighting caused by the laser as the scene changes, generating additional IP traffic. 
The excessive traffic traverses the network to reach the camera viewing station. 
Since the traffic must pass through a physical resource shared with other network devices, overloaded shared resources can lead to congestion and packet drops. 
Consequently, if the network is not properly physically segmented, this attack potentially affects critical network segments.
The more cameras the attacker can see (verbatim), the greater the potential effect of the DoS.
In a wired setting, this physical resource would typically be a shared link, such as the downstream links between network switches. In a wireless setting, on the other hand, this shared resource would be the transmission queue of the wireless access point.

\subsection{Assembly of a Laser Flicker}
\label{sec:our_attack}

To increase an IP camera's bitrate to the extent required to ensure the attack's success, we should generate a signal that modern video encodings cannot efficiently compress. 
Algorithms used for this purpose perform many optimizations to reduce the video's bitrate to the greatest extent possible without harming quality (see \autoref{sec:compression}). 
However, rapid unnatural changes in the scene reduce the compression quality to a great extent. 
To generate such rapid changes an attacker located at the direct line of sight from a camera, can dazzle it using a flickering laser beam.

\begin{figure}[!htb]
\centering
\begin{tcolorbox}[
   colback=gray!5,
   colframe=gray!40!black,
   width=0.95\columnwidth,
   arc=5mm,
   boxrule=0.5pt
]
\begin{subfigure}[b]{\columnwidth}
   \centering
   \begin{tcolorbox}[
       colback=blue!3,
       colframe=blue!30!white,
       arc=2mm,
       boxrule=0.3pt,
       width=0.95\columnwidth
   ]
    \includegraphics[width=\columnwidth]{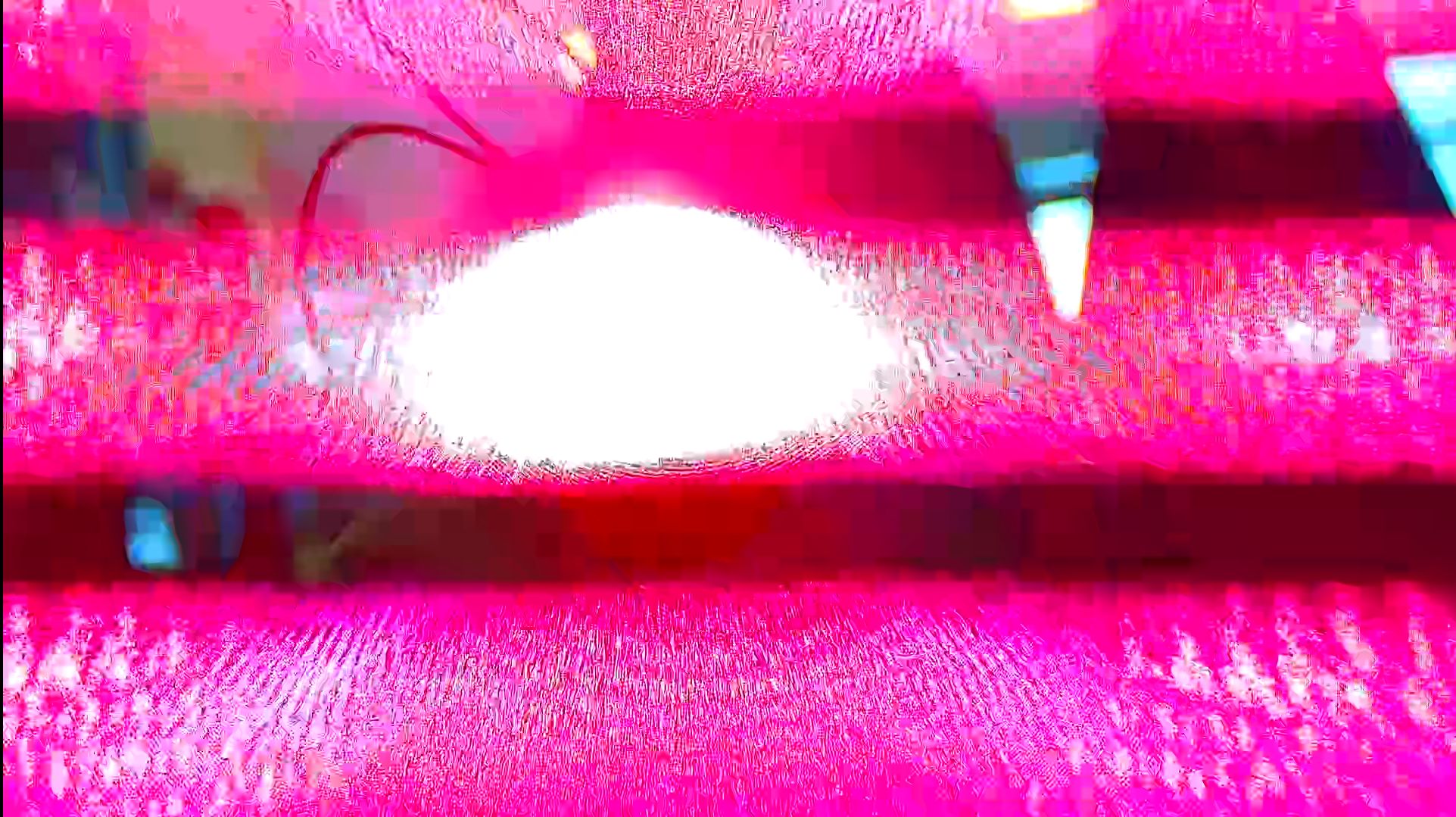}
    \caption{Horizontal $0^\circ$, vertical $0^\circ$}
   \label{fig:dazzling-0}
   \end{tcolorbox}
\end{subfigure}

\vspace{2mm}

\begin{subfigure}[b]{\columnwidth}
   \centering
   \begin{tcolorbox}[
       colback=blue!3,
       colframe=blue!30!white,
       arc=2mm,
       boxrule=0.3pt,
       width=0.95\columnwidth
   ]
    \includegraphics[width=\columnwidth]{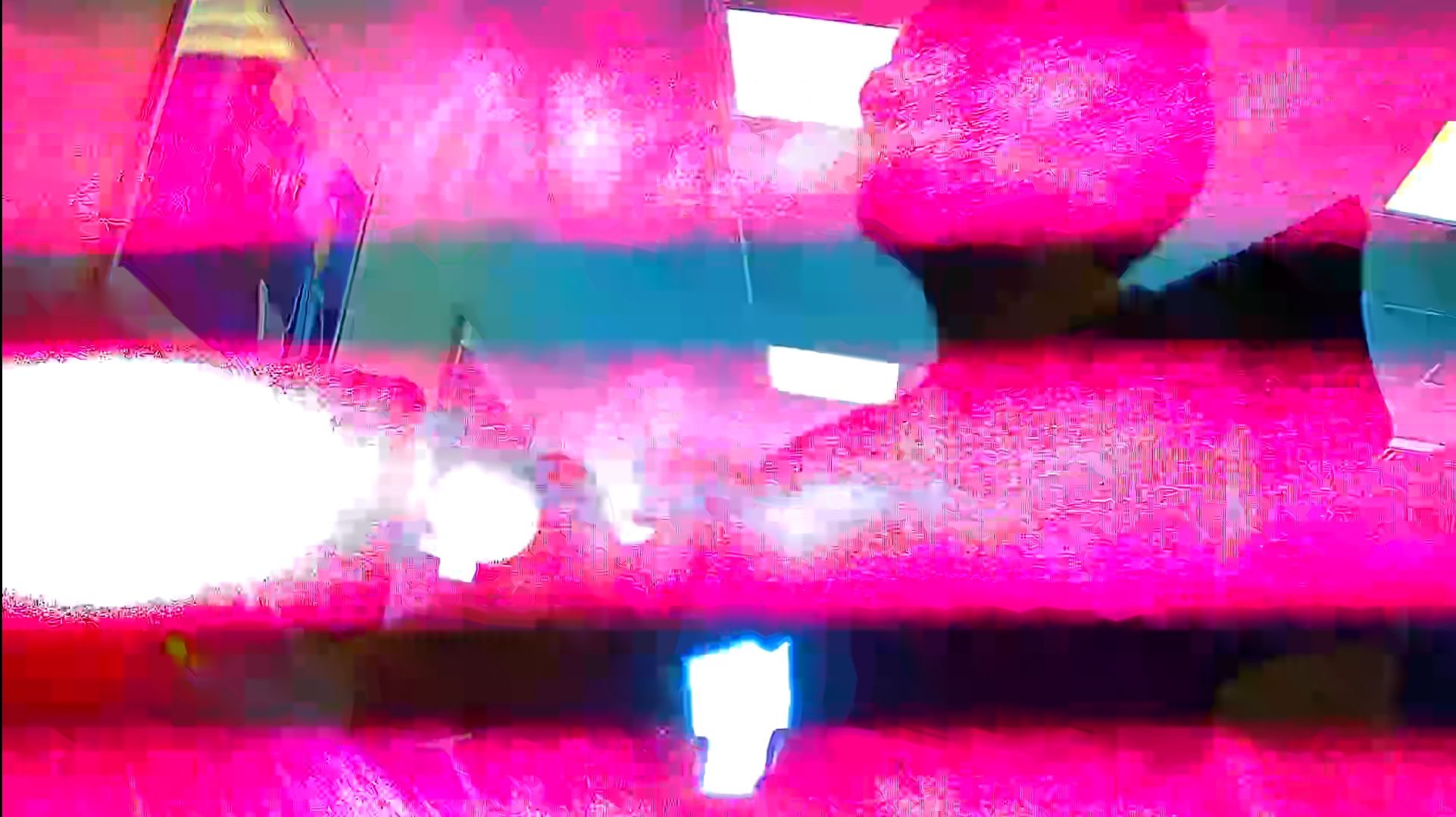}
    \caption{Horizontal $30^\circ$, vertical $0^\circ$}
   \label{fig:dazzling-30-0}
   \end{tcolorbox}
\end{subfigure}

\vspace{2mm}

\begin{subfigure}[b]{\columnwidth}
   \centering
   \begin{tcolorbox}[
       colback=blue!3,
       colframe=blue!30!white,
       arc=2mm,
       boxrule=0.3pt,
       width=0.95\columnwidth
   ]
    \includegraphics[width=\columnwidth]{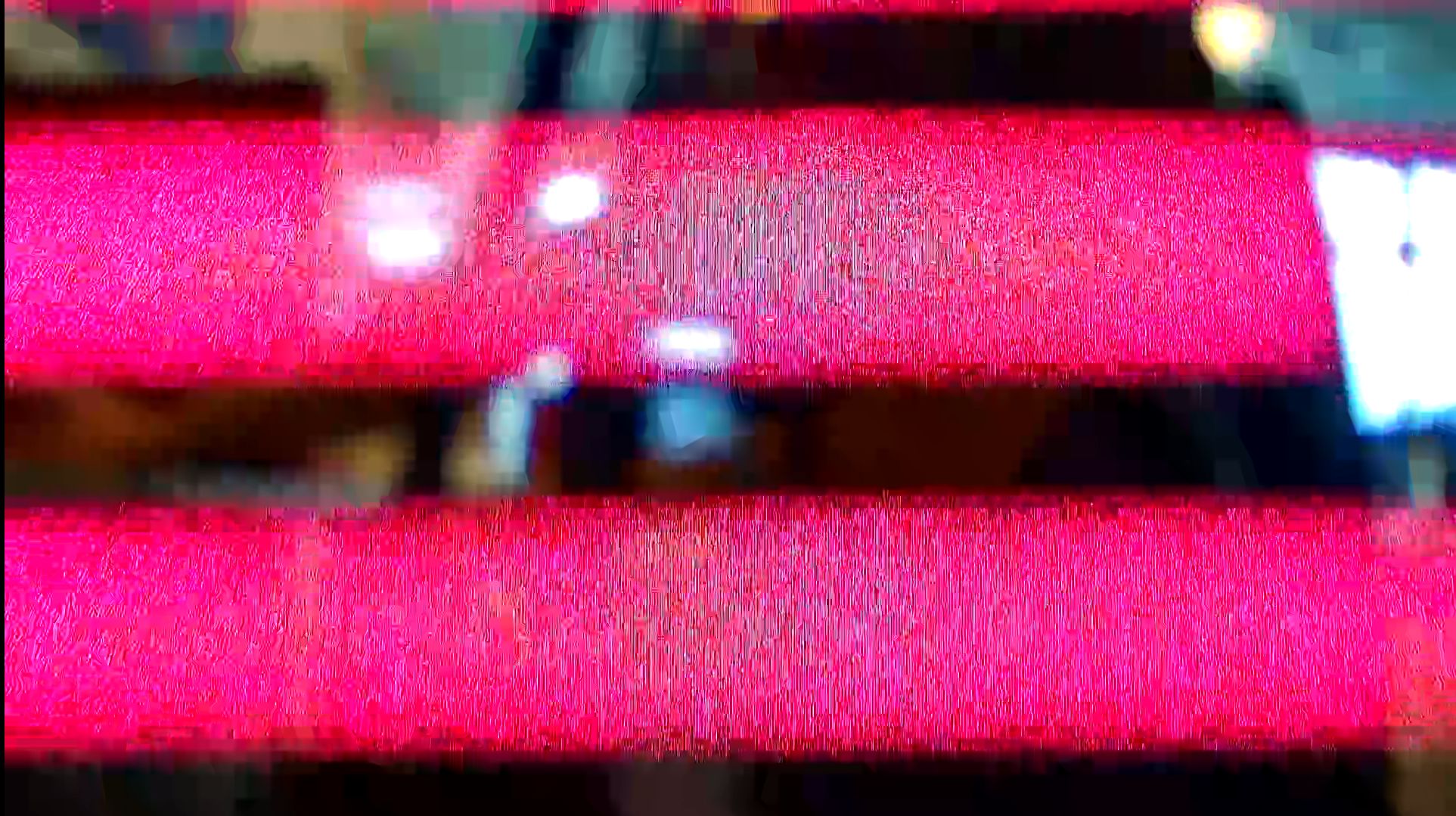}
    \caption{Horizontal $0^\circ$, vertical $40^\circ$}
   \label{fig:dazzling-0-40}
   \end{tcolorbox}
\end{subfigure}

\end{tcolorbox}

\caption{The effect of camera dazzling from different incidence angles. The horizontal stripes are highly dynamic. }
\label{fig:dazzling}
\end{figure}

As detailed in \cite{Booth2018LaserDazzling}, dazzling an IP camera with a laser has many physical effects, including light diffraction, reflections, and scattering of the laser beam. 
Some of these effects are sensitive to changes in the laser beam's position. 
Even when a laser beam points directly into the camera lens, the incidence angle affects how the beam reflects as depicted in \autoref{fig:dazzling}.
We explore the effect of laser positioning on overall traffic generated by the camera in \autoref{sec:evaluation:positioning}.
When the laser is well-positioned, these effects can take over most or all of the camera's view,  
in most cases, yielding a frame with little resemblance to the original scene (see discussion in \autoref{sec:discussion}).

\begin{figure}[!htbp]
    \centering
    \includegraphics[width=0.7\columnwidth]{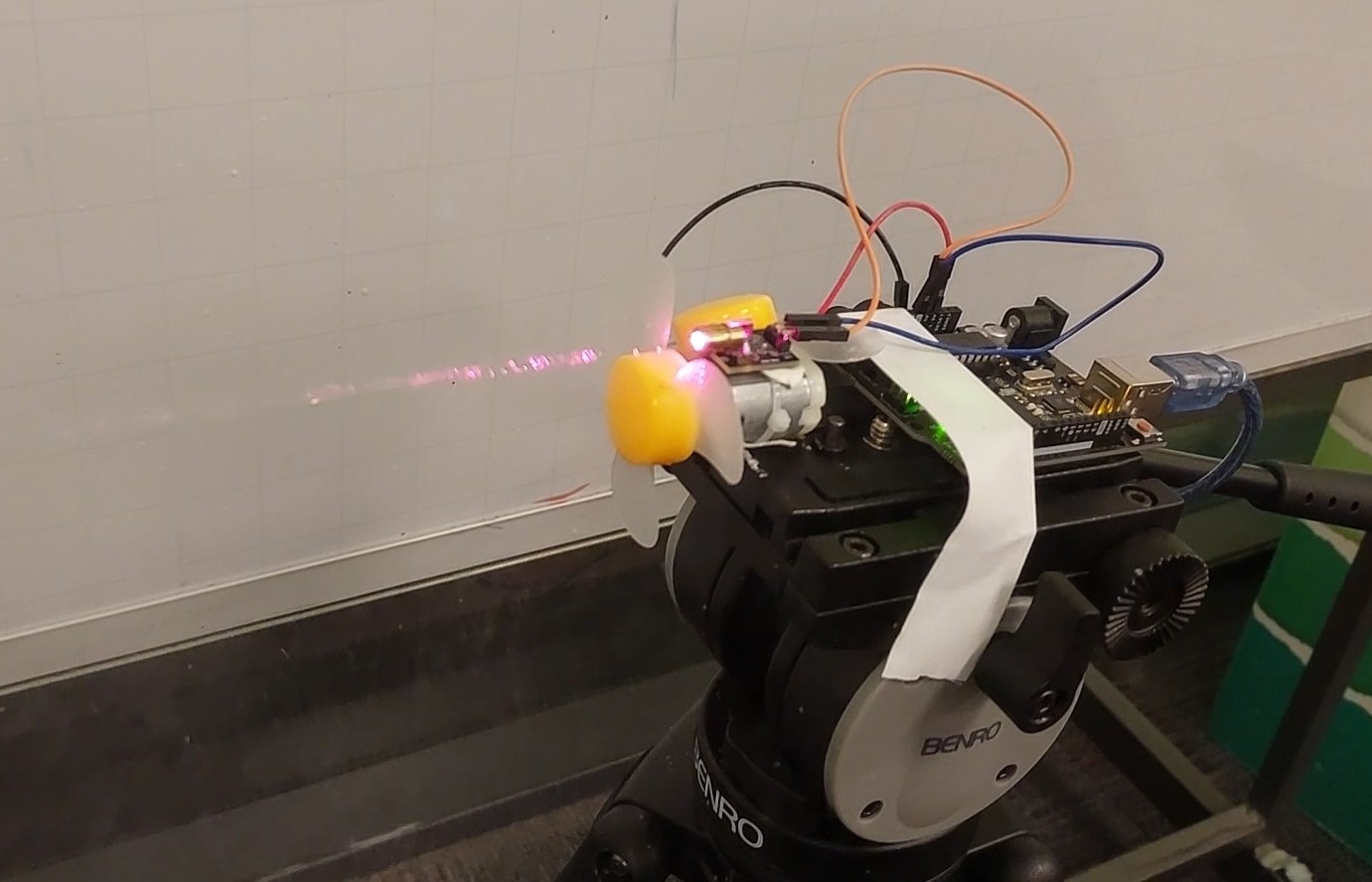}
    \caption{Laser mount}
    \label{fig:fan}
\end{figure}

To assemble the attack device, we used a simple LED laser diode.
The laser is controlled by an Arduino. 
To fix the laser position and direction, we mounted it on a camera tripod commonly used by photographers (see \autoref{fig:fan}).
Although the Arduino can turn the laser on and off, the frequency of doing so is not high enough. 
A fan in front of the laser can generate very high frequencies by blocking the beam with its wings.  
The horizontal stripes in \autoref{fig:dazzling} are due to the combined effect of the flickering frequency and the area scan algorithm of high-resolution cameras.

Using a fan to create the flicker has an additional (not-so) surprising effect that increases the video streaming bitrate.   
A small servomotor powering fan mounted on the tripod in front of the laser emitter generates small vibrations that significantly affect the light reflections and scattering.
The small vibrations of the motor create slight movements without causing the laser to lose focus on its target.
These vibrations introduce additional chaotic behavior to the captured frames further increasing the frame-to-frame differences and hence the bitrate.  

In \autoref{tab:AttackCompression}, we summarize the effect of rapid changes triggered by the laser flicker on the camera's video stream optimizations.
Overall, using the laser flicker, we achieved a bitrate increase of around $6 \times$, up from 2.25 Mb/s to 13.5 Mb/s as elaborated in the next section.

\begin{table*}[!htb]
\centering
\begin{tcolorbox}[
    colback=gray!5,
    colframe=gray!40!black,
    width=\textwidth,
    arc=5mm,
    boxrule=0.5pt,
    center,
    halign=center
]
\footnotesize
\begin{tabularx}{\textwidth}{X|X}
    \toprule
    \textbf{Video Compression Optimization} & \textbf{Laser Flickering} \\
    \midrule
    Color information is compressed, while dark-bright contrasts are not. & Laser diffraction yields very uniform colors, as well as many dark-bright contrasts, which are further exacerbated by the laser's flickering. \\
    \midrule
    Recent past (and future) frames are used to encode the current one. Specifically, the movement of an object from one location to another is encoded as a motion vector rather than as a retransmission of the object. & Frames vary wildly from one to the other, and these variations cannot be described or even approximated as linear movement. \\
       \bottomrule
\end{tabularx}
\end{tcolorbox}
\caption{Effect of Laser Flickering on Video Compression}
\label{tab:AttackCompression}
\end{table*}

\section{Evaluation}
\label{sec:evaluation}

In this section, we describe the experimental setup, the metrics used to measure the performance of the attack, and the results obtained in different settings.

\subsection{Testbed}

To evaluate the effectiveness of the attack in various settings, we constructed both a wired and wireless testbed. We built both our setups around a Sunba Performance Series camera and a KY-008 laser transmitter module aimed directly at the camera's lens. The camera was cable-connected to the network, while the rest of the network can be wired or wireless. 

We benchmarked both wired and wireless setups with networking hardware made by different manufacturers, to ensure that the attack's performance is not simply due to one manufacturer's implementation decisions.


\subsubsection{Wired Setup} \label{sec:wired_setup}

\begin{figure}[!htbp]
    \centering
    \includegraphics[width=\columnwidth]{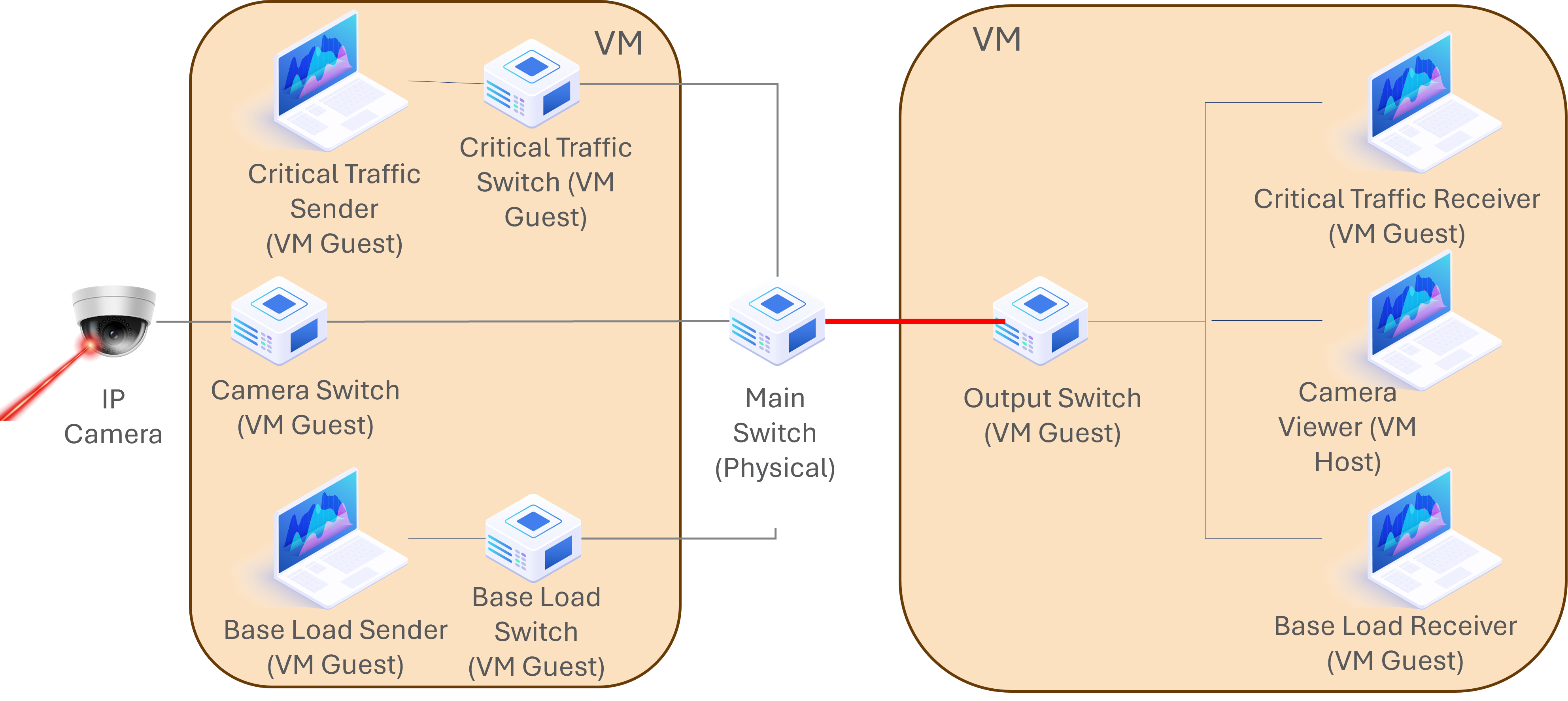}
    \caption{Wired testbed topology.}
    \label{fig:WiredTopology}
\end{figure}

Our wired testbed setup is presented in \autoref{fig:WiredTopology} and consists of the following components: 
     \newline\noindent\textbf{Main switch} (we used both Cisco Catalyst 2960-C and a Juniper SSG-5 with firewall capabilities in our evaluation, both of which yielded similar results) with a 100Mb/s port that we wish to overload; 
    \newline\noindent\textbf{IP Camera} connected to a virtual switch (implemented as a virtual machine (VM) guest) via Ethernet to USB adapter. The traffic from the camera traverses the network to reach the camera viewer host;
    \newline\noindent\textbf{Critical traffic sender and critical traffic receiver} implemented as VMs (used to emulate time-sensitive network traffic sent between two locations on the network);
    \newline\noindent\textbf{Base-load sender and base-load receiver} implemented as VMs (used to generate both a constant network load that simulates baseline activity on the network and different network load profiles for network disruption testing, as elaborated on below).
    
As can be seen in \autoref{fig:WiredTopology}, some of the components are physical, and some are virtual. Emulation was performed using VirtualBox on a computer with 32 GB of RAM running Lubuntu 24.04.1. Each VM ran Ubuntu Desktop 24.04.1 and was assigned four threads and 2 or 4 GB of RAM depending on its need, ensuring neither caused a bottleneck. 


\begin{figure}[!htbp]
    \centering
    \includegraphics[width=\columnwidth]{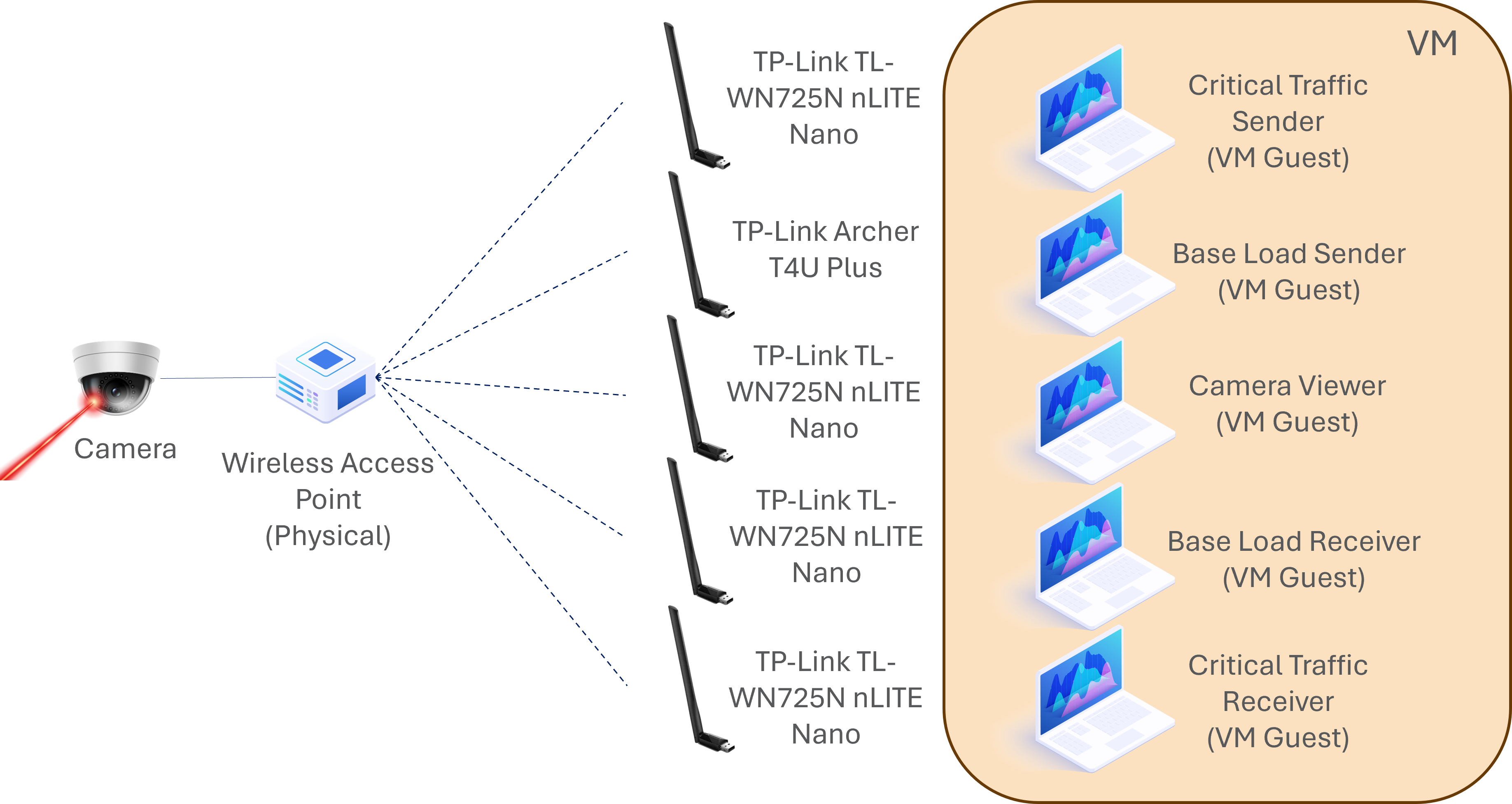}
    \caption{Wireless testbed topology.}
    \label{fig:WirelessTopology}
\end{figure}

\subsubsection{Wireless Setup} \label{sec:wireless_setup}

Our wireless testbed setup is presented in \autoref{fig:WirelessTopology}, and it consists of the same end devices serving the same roles as in the wired testbed; this time, however, all of the devices are wirelessly connected to a wireless access point as follows:
\newline\noindent\textbf{Wireless access point} (we used TP-Link, TL-WA901ND, TP-Link TL-WR840N, and a D-Link DIR-825, all of which yielded similar results) that we wish to overload.
    \newline\noindent\textbf{The IP camera} is cable-connected to an Ethernet port of the access point.
    \newline\noindent\textbf{Wireless network adapters} TP-Link TL-WN725N nLITE Nano were chosen to model a highly loaded network due to their slow transmission rates. They are used by the camera viewer, critical traffic sender, critical traffic receiver, and base-load receiver. The base-load sender uses a standard TP-Link Archer T4U Plus network adapter, as it needs to be able to transmit constant network loads to stress the network.
    
As devices connected to a wireless system are typically end devices, we omitted the virtualized switches from our final testbed, but the attack should also be effective with such intermediary devices.

\subsection{Evaluation of Network Disruption}

\subsubsection{Controlled Evaluation Parameters}

We performed the evaluation with six \textbf{base-load} profiles, each creating load by transmitting 64KB UDP payloads using hping3.

In the wired setting, we used base-loads of either 83.2 Mb/s; 86.4 Mb/s; or 89.6 Mb/s sent between the base-load sender and base-load receiver. 
In the wireless setting, we used base-loads of 22.4 Mb/s; 25.6 Mb/s; or 28.8 Mb/s.


\subsubsection{Evaluation Metrics} \label{sec:disruption}

To evaluate the effectiveness of the attack, we created a list of disruption metrics. Since different networks may have different traffic types as their baseline, we wanted to make sure that the metrics created would apply to various common network scenarios, ensuring that our estimations would be as relevant as possible.
Accordingly, our disruption metrics take into account both the UDP and TCP to estimate how different applications may be affected. We also use ICMP, and perform file transfers in our evaluation. We account for the packet drop rate over different sizes and protocols, to obtain an estimate of the number of retransmissions required in different scenarios. We also use benchmarks that retransmit dropped or timed-out packets, and measure the difference in data reliably transmitted over time. A list of our disruption metrics is provided in \autoref{tab:Benchmarks}.

\begin{table*}[!hbt]  
\begin{tcolorbox}[
    colback=gray!5,
    colframe=gray!40!black,
    width=\textwidth,  
    arc=5mm,
    boxrule=0.5pt,
    center,
    halign=center
]
\footnotesize
    \centering
    \begin{tabularx}{\linewidth}{  
        >{\raggedright\arraybackslash}X
        >{\raggedright\arraybackslash}X
        >{\raggedright\arraybackslash}X
        >{\raggedright\arraybackslash}X
    }
    \toprule
    \textbf{Name of Metric} & \textbf{Tool Used} & \textbf{Description} & \textbf{Purpose} \\
    \midrule
    Roundtrip ICMP &
    hping3 (ICMP) &
    Transmit ICMP echo requests and receive echo replies back &
    Test dropped pings during data exchange \\
    \midrule
    One-way TCP data &
    hping3 (TCP) &
    Transmit TCP data and receive an ACK back &
    Test dropped pings for one-way data transfer \\
    \midrule
    File transfer &
    rsync (SSH) &
    Transmit a 10\,MB file over SSH &
    Real-world network bandwidth test \\
    \midrule
    TCP stress test &
    iperf3 (TCP) &
    Attempt to transmit 100\,Mb/s TCP traffic for 60 seconds &
    Measure bandwidth with congestion control \\
    \midrule
    UDP stress test &
    iperf3 (UDP) &
    Attempt to transmit 100\,Mb/s UDP traffic for 60 seconds &
    Measure bandwidth without congestion control \\
    \bottomrule
    \end{tabularx}
    \end{tcolorbox}
    \caption{Network Disruption Metrics}
    \label{tab:Benchmarks}
\end{table*}

\subsubsection{Measuring Impact of Laser Positioning}
\label{sec:evaluation:positioning}

\begin{figure}[!htb]
\centering
\begin{tcolorbox}[
   colback=gray!5,
   colframe=gray!40!black,
   width=0.95\columnwidth,
   arc=5mm,
   boxrule=0.5pt
]
\begin{subfigure}[b]{\columnwidth}
   \centering
   \begin{tcolorbox}[
       colback=blue!3,
       colframe=blue!30!white,
       arc=2mm,
       boxrule=0.3pt,
       width=0.95\columnwidth
   ]
   \includegraphics[width=\columnwidth]{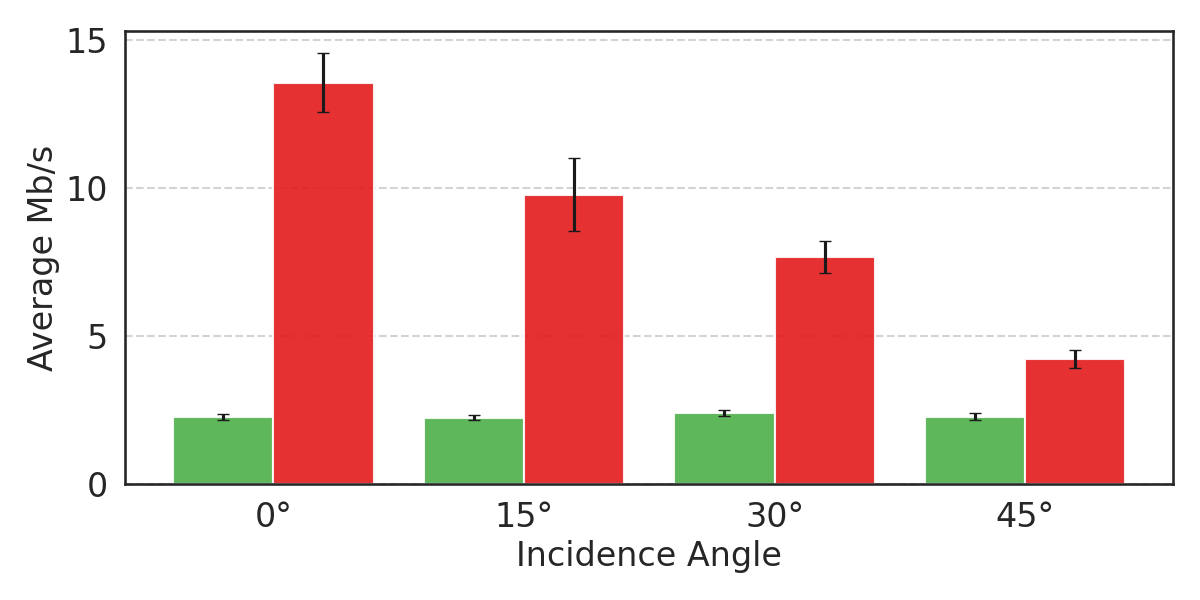}
   \caption{Traffic volume as a function of the horizontal incidence angle}
   \label{fig:LeftRightPositioning}
   \end{tcolorbox}
\end{subfigure}

\vspace{2mm}

\begin{subfigure}[b]{\columnwidth}
   \centering
   \begin{tcolorbox}[
       colback=blue!3,
       colframe=blue!30!white,
       arc=2mm,
       boxrule=0.3pt,
       width=0.95\columnwidth
   ]
   \includegraphics[width=\columnwidth]{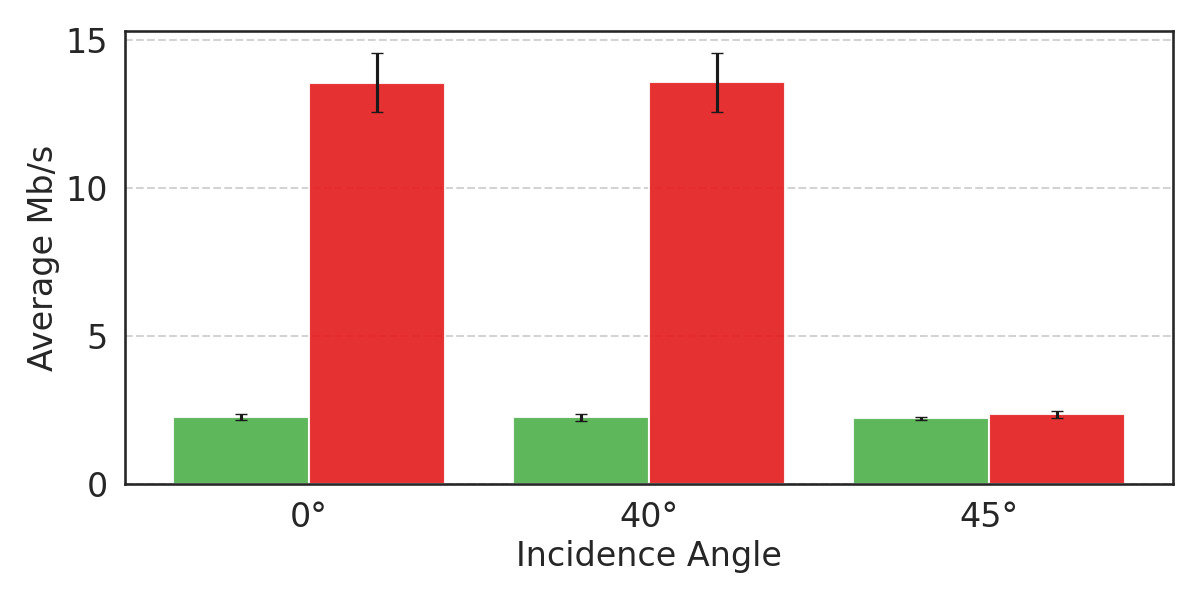}
   \caption{Traffic volume as a function of the vertical incidence angle}
   \label{fig:UpDownPositioning}
   \end{tcolorbox}
\end{subfigure}
\end{tcolorbox}

\vspace{-2mm}
\begin{center}
   \includegraphics[scale=0.2]{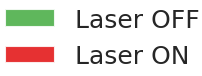}
\end{center}
\vspace{-5mm}
\caption{Impact of laser beam incidence angle}
\label{fig:LaserPositioning}
\end{figure}

The attack was also tested in settings where the laser was at various angles in relation to the camera. When changing the horizontal angle, we have found the increase in bitrate diminishes as the angle increases, as can be seen in \autoref{fig:LeftRightPositioning}. However, when changing the vertical angle, the bitrate remains fairly constant before reaching an abrupt cutoff at \textasciitilde 45\textdegree, as can be seen in \autoref{fig:UpDownPositioning}.

\subsubsection{Wired Attack Results}

\begin{figure*}[!htb]
\centering
\begin{tcolorbox}[
   colback=gray!5,
   colframe=gray!40!black,
   width=\textwidth,
   arc=5mm,
   boxrule=0.5pt
]
\begin{subfigure}[b]{0.325\textwidth}
   \centering
   \begin{tcolorbox}[
       colback=blue!3,
       colframe=blue!30!white,
       arc=2mm,
       boxrule=0.3pt
   ]
   \includegraphics[width=\textwidth]{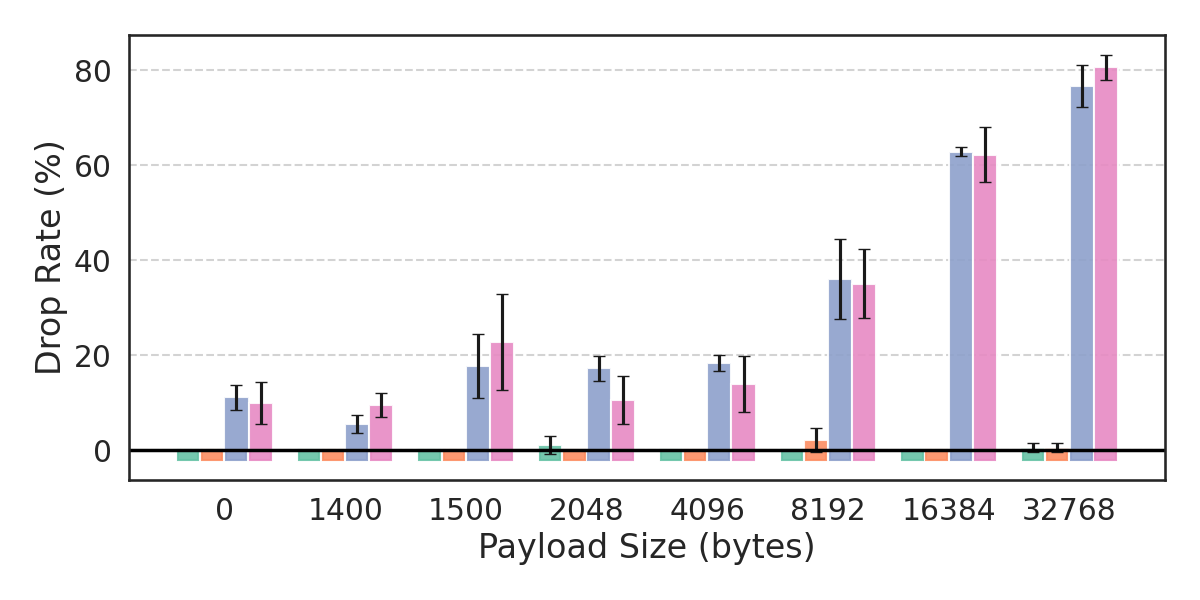}
   \caption{Drop rate at 83.2 Mb/s base-load.}
   \label{fig:drop1}
   \end{tcolorbox}
\end{subfigure}
\hfill
\begin{subfigure}[b]{0.325\textwidth}
   \centering
   \begin{tcolorbox}[
       colback=blue!3,
       colframe=blue!30!white,
       arc=2mm,
       boxrule=0.3pt
   ]
   \includegraphics[width=\textwidth]{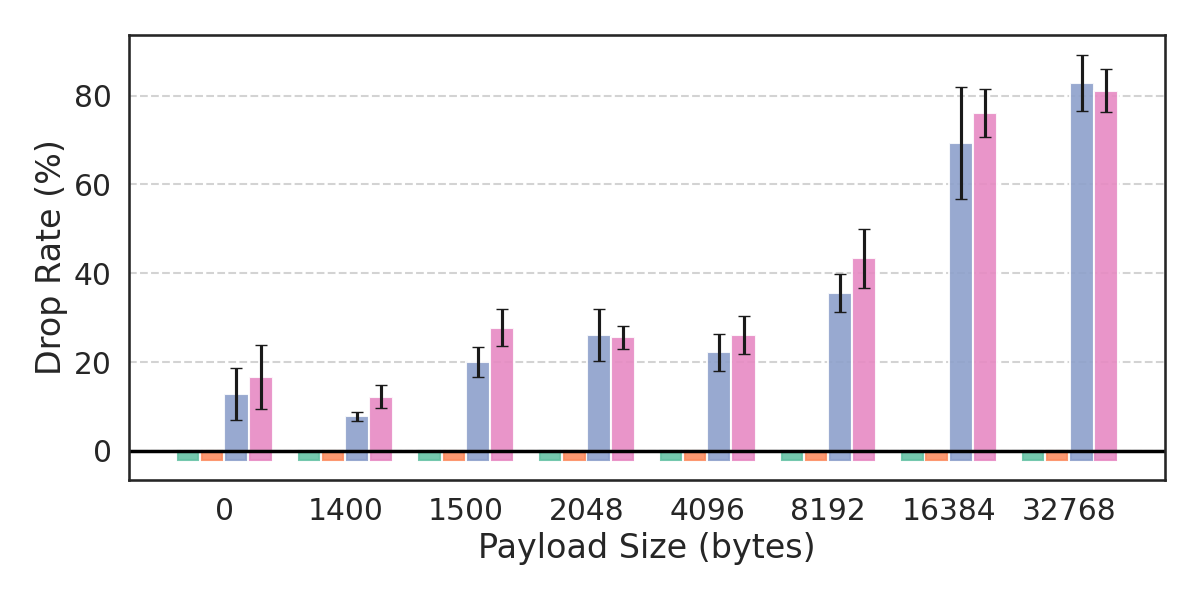}
   \caption{Drop rate at 86.4 Mb/s baseload.}
   \label{fig:drop2}
   \end{tcolorbox}
\end{subfigure}
\hfill
\begin{subfigure}[b]{0.325\textwidth}
   \centering
   \begin{tcolorbox}[
       colback=blue!3,
       colframe=blue!30!white,
       arc=2mm,
       boxrule=0.3pt
   ]
   \includegraphics[width=\textwidth]{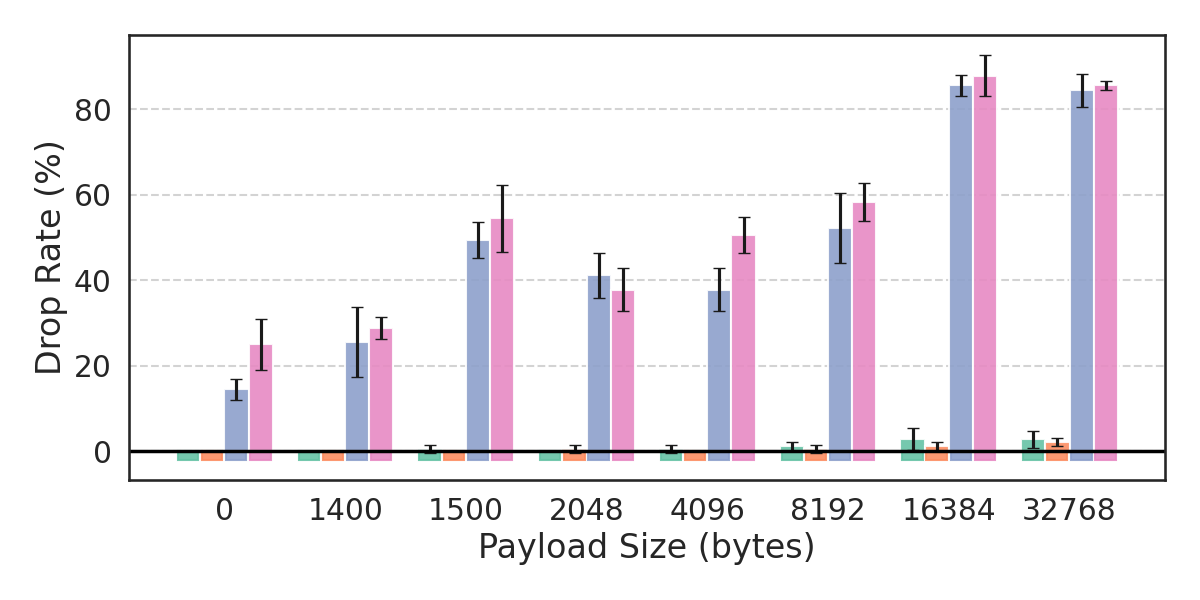}
   \caption{Drop rate at 89.6 Mb/s base-load.}
   \label{fig:drop3}
   \end{tcolorbox}
\end{subfigure}

\vspace{2mm}

\begin{subfigure}[b]{0.325\textwidth}
   \centering
   \begin{tcolorbox}[
       colback=blue!3,
       colframe=blue!30!white,
       arc=2mm,
       boxrule=0.3pt
   ]
   \includegraphics[width=\textwidth]{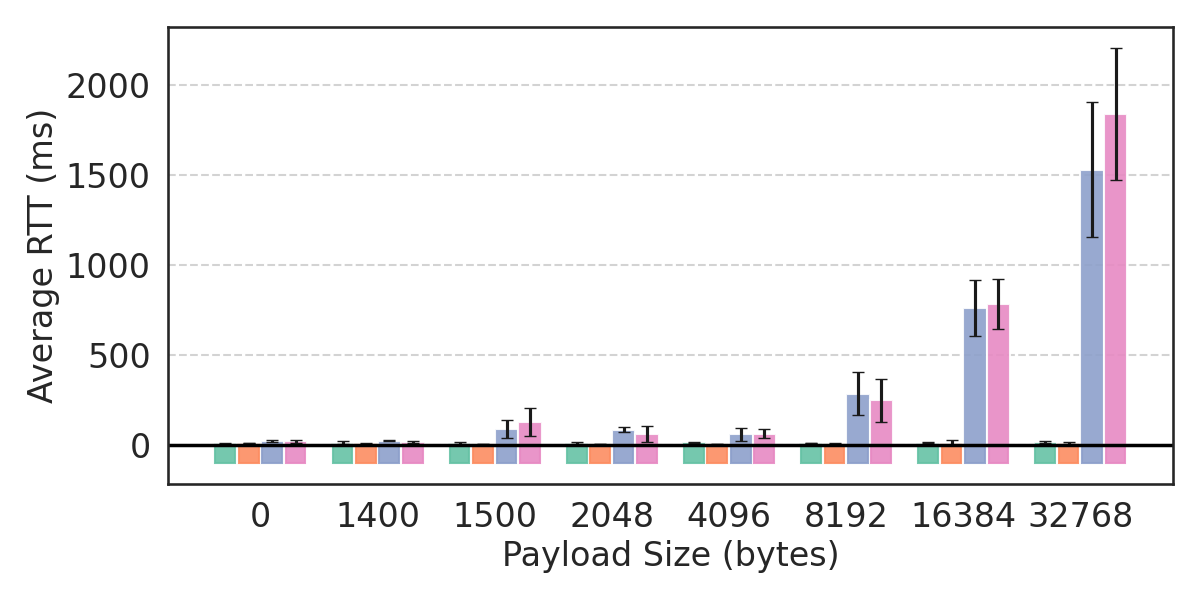}
   \caption{Average RTT at 83.2 Mb/s base-load.}
   \label{fig:rtt1}
   \end{tcolorbox}
\end{subfigure}
\hfill
\begin{subfigure}[b]{0.325\textwidth}
   \centering
   \begin{tcolorbox}[
       colback=blue!3,
       colframe=blue!30!white,
       arc=2mm,
       boxrule=0.3pt
   ]
   \includegraphics[width=\textwidth]{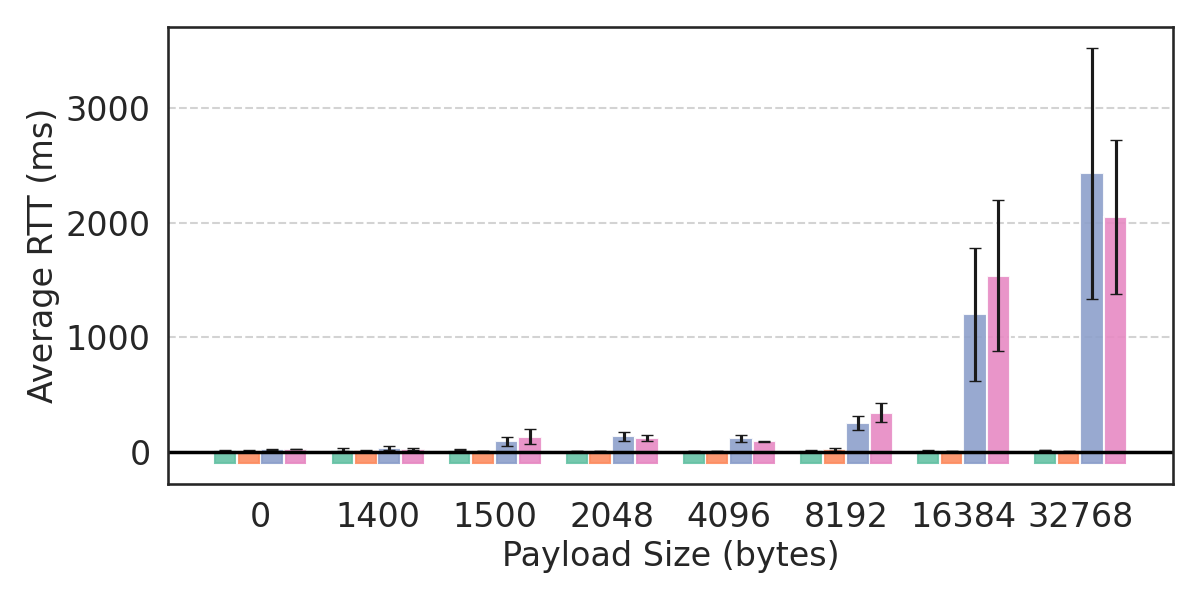}
   \caption{Average RTT at 86.4 Mb/s base-load.}
   \label{fig:rtt2}
   \end{tcolorbox}
\end{subfigure}
\hfill
\begin{subfigure}[b]{0.325\textwidth}
   \centering
   \begin{tcolorbox}[
       colback=blue!3,
       colframe=blue!30!white,
       arc=2mm,
       boxrule=0.3pt
   ]
   \includegraphics[width=\textwidth]{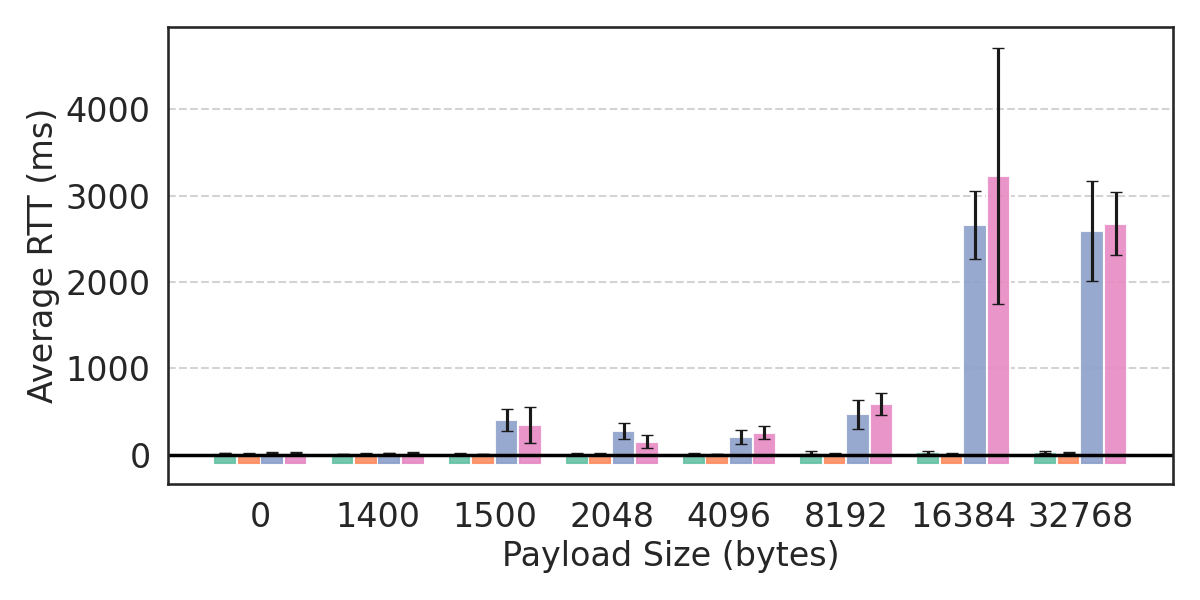}
   \caption{Average RTT at 89.6 Mb/s base-load.}
   \label{fig:rtt3}
   \end{tcolorbox}
\end{subfigure}
\end{tcolorbox}

\vspace{-3mm}
\begin{center}
   \includegraphics[scale=0.2]{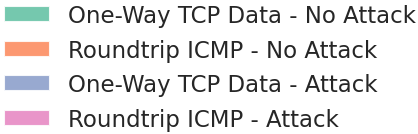}
\end{center}
\vspace{-3mm}
\caption{Roundtrip ICMP and one-way TCP data in wired setup}
\label{fig:WiredEval}
\end{figure*}

The results of our wired attack on the Cisco Catalyst 2960-C switch are presented in \autoref{fig:WiredEval}; as can be seen, the attack caused significant network disruptions. The figure shows how roundtrip ICMP and one-way TCP data disruption metrics change once the attack is executed, under different base-load profiles.
Baseline performance for the roundtrip ICMP and one-way TCP data metrics records roundtrip times (RTTs) below 30 ms with nearly no packet loss under normal conditions. During the attack, RTTs increased dramatically, reaching up to three seconds, along with a packet drop rate of 10–20\% under optimal conditions and up to 80\% in more demanding settings. These results were consistent across TCP and ICMP traffic due to the full-duplex data transmission of the examined Cisco Catalyst 2960-C switch.

More specifically, in the upper half of the figure, we can see how the drop rate in both disruption metrics increases along with the increase of base-load and the payload size generated under these tests. 
For example, the pink bars show how the ICMP echo requests and replies (i.e., Roundtrip ICMP disruption metric) are affected when the camera is dazzled with a laser, as opposed to the orange bars, which show how this traffic would normally behave when there is no attack in progress. 
As we can see, the orange bars remain very low even as the payload size of the ICMP traffic increases (as can be seen in the same figure), and as the base load is increasing (as can be seen between figures). 
Conversely, once the attack kicks-in, this traffic starts suffering from increasing packet loss (as shown by the pink bars), which grows as the payload size increases (as can be seen in the same figure), and as the base load is increasing (as can be seen between figures).

The same is true for tests in which we recorded the average RTT of traffic, as can be seen in the lower half of \autoref{fig:WiredEval}. For example, in attempts to transmit TCP data and receive an ACK back (i.e., One-Way TCP Data disruption metric)--as reflected by green (no attack in place) and blue (attack in progress) bars in the figure--we have seen a significant increase in RTT once the attack was in progress, as the payload size of the TCP traffic and the base load increased. This was in contrast to the baseline recorded when there was no attack in progress, where the RTT remained relatively constant, regardless of the increase in payload size and base-load. 

\begin{figure}[!htb]
\begin{tcolorbox}[
    colback=gray!5,
    colframe=gray!40!black,
    width=\columnwidth,
    arc=5mm,
    boxrule=0.5pt
]
\centering
    \begin{subfigure}[b]{\columnwidth}
        \centering
        \includegraphics[width=\columnwidth]{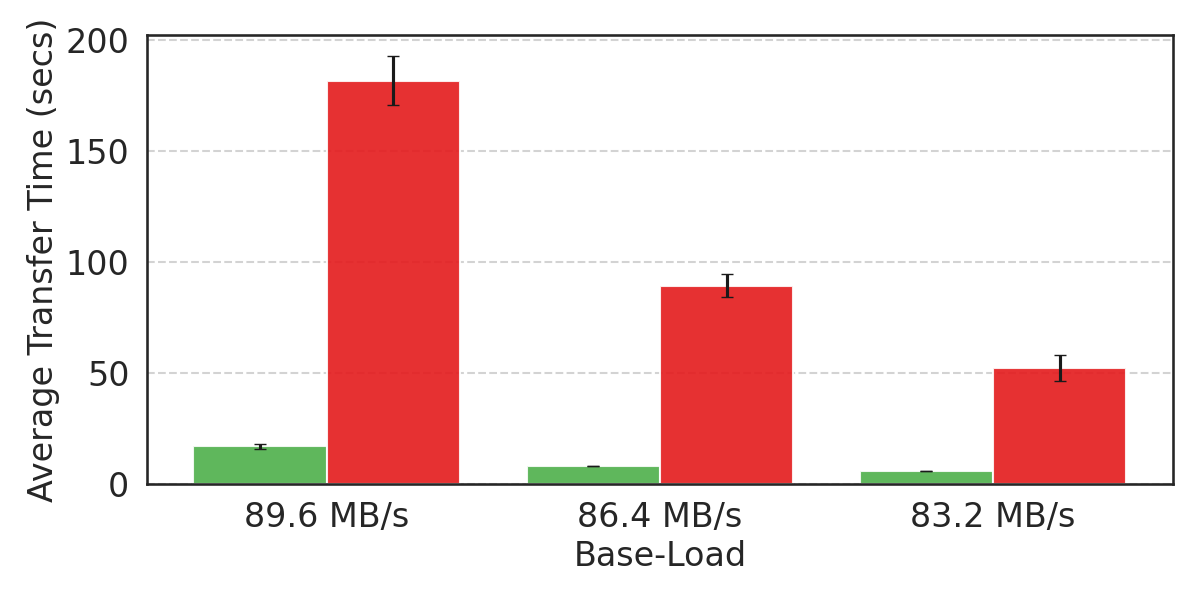}
         \label{fig:subfig2}
    \end{subfigure}
    \end{tcolorbox}
    \vspace{-10pt}  
\begin{center}
        \includegraphics[scale=0.2]{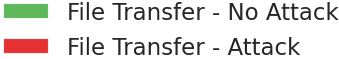}
    \end{center}
\vspace{-3mm}
\caption{File transfer in wired setup}
\label{fig:rsync}
\end{figure}

Similarly, the file transfer and TCP stress test metric--presented in \autoref{fig:iperf} and \autoref{fig:rsync}--showed up to a 90\% reduction in effective throughput, largely due to congestion control mechanisms, while the UDP stress test metric showed a 5–10\% decrease. This is proportional to the video stream's share of overall traffic and reflects the protocol’s lack of congestion management.

Overall, the wired setup results confirm that the attack greatly impacts network reliability and efficiency by overwhelming critical network pathways. We note that as the camera begins generating additional traffic as soon as the laser is directed at it, the results described above occur near-instantly.

\subsubsection{Wireless Attack Results}
\begin{figure*}[!htb]
\centering
\begin{tcolorbox}[
   colback=gray!5,
   colframe=gray!40!black,
   width=\textwidth,
   arc=5mm,
   boxrule=0.5pt
]
\begin{subfigure}[b]{0.325\textwidth}
   \centering
   \begin{tcolorbox}[
       colback=blue!3,
       colframe=blue!30!white,
       arc=2mm,
       boxrule=0.3pt
   ]
   \includegraphics[width=\textwidth]{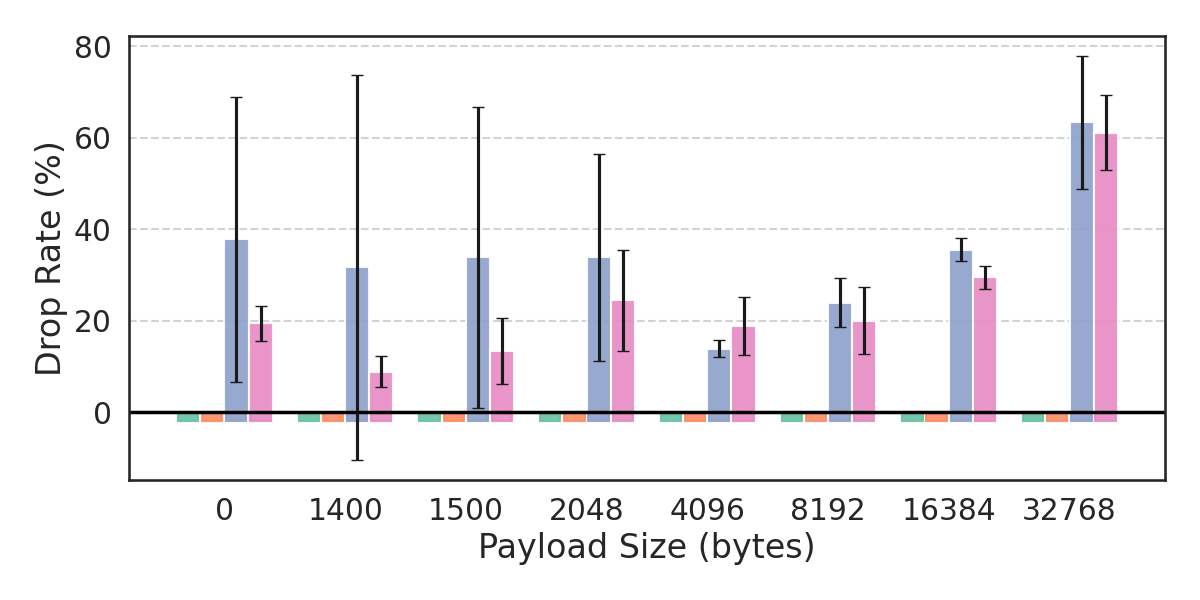}
   \caption{Drop rate at 22.4 Mb/s base-load.}
   \label{fig:wireless_drop1}
   \end{tcolorbox}
\end{subfigure}
\hfill
\begin{subfigure}[b]{0.325\textwidth}
   \centering
   \begin{tcolorbox}[
       colback=blue!3,
       colframe=blue!30!white,
       arc=2mm,
       boxrule=0.3pt
   ]
   \includegraphics[width=\textwidth]{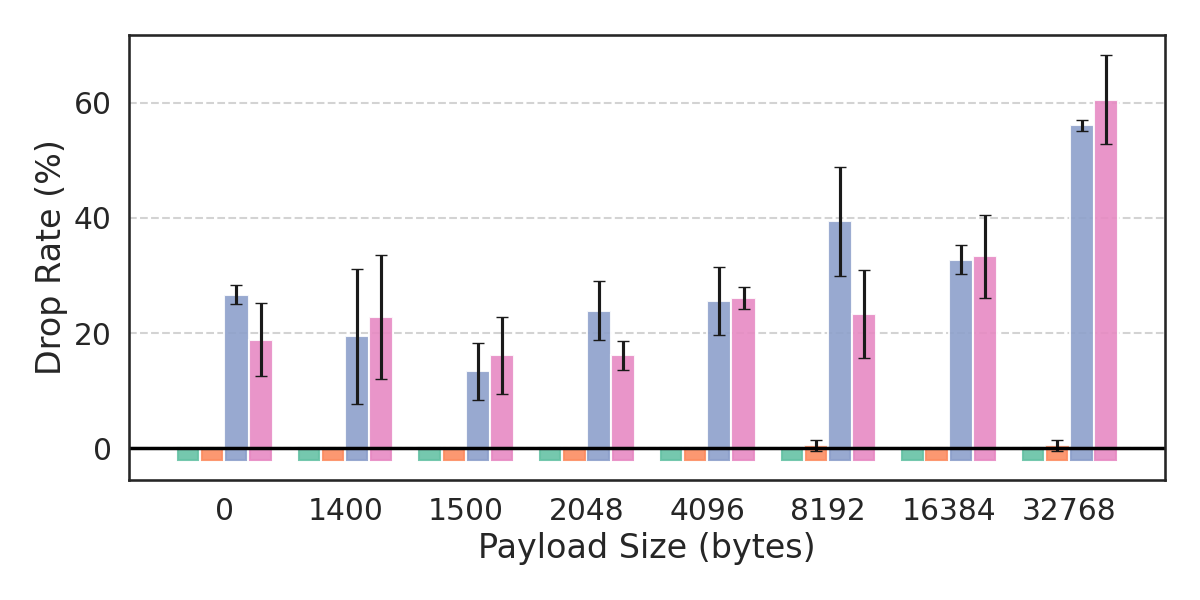}
   \caption{Drop rate at 25.6 Mb/s base-load.}
   \label{fig:wireless_drop2}
   \end{tcolorbox}
\end{subfigure}
\hfill
\begin{subfigure}[b]{0.325\textwidth}
   \centering
   \begin{tcolorbox}[
       colback=blue!3,
       colframe=blue!30!white,
       arc=2mm,
       boxrule=0.3pt
   ]
   \includegraphics[width=\textwidth]{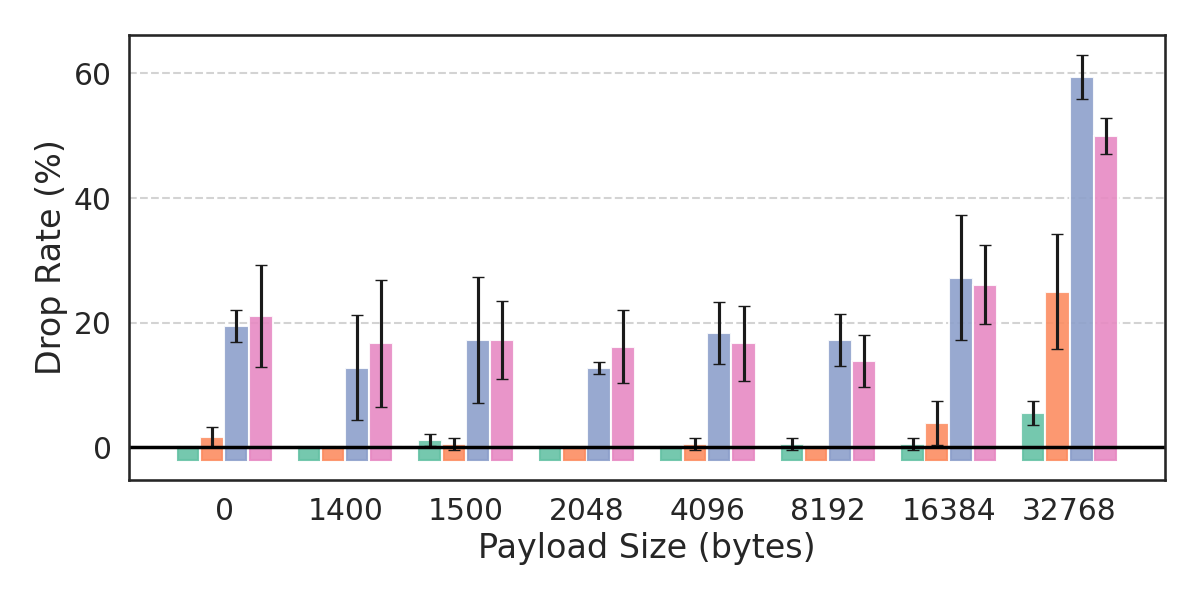}
   \caption{Drop rate at 28.8 Mb/s base-load.}
   \label{fig:wireless_drop3}
   \end{tcolorbox}
\end{subfigure}

\vspace{2mm}

\begin{subfigure}[b]{0.325\textwidth}
   \centering
   \begin{tcolorbox}[
       colback=blue!3,
       colframe=blue!30!white,
       arc=2mm,
       boxrule=0.3pt
   ]
   \includegraphics[width=\textwidth]{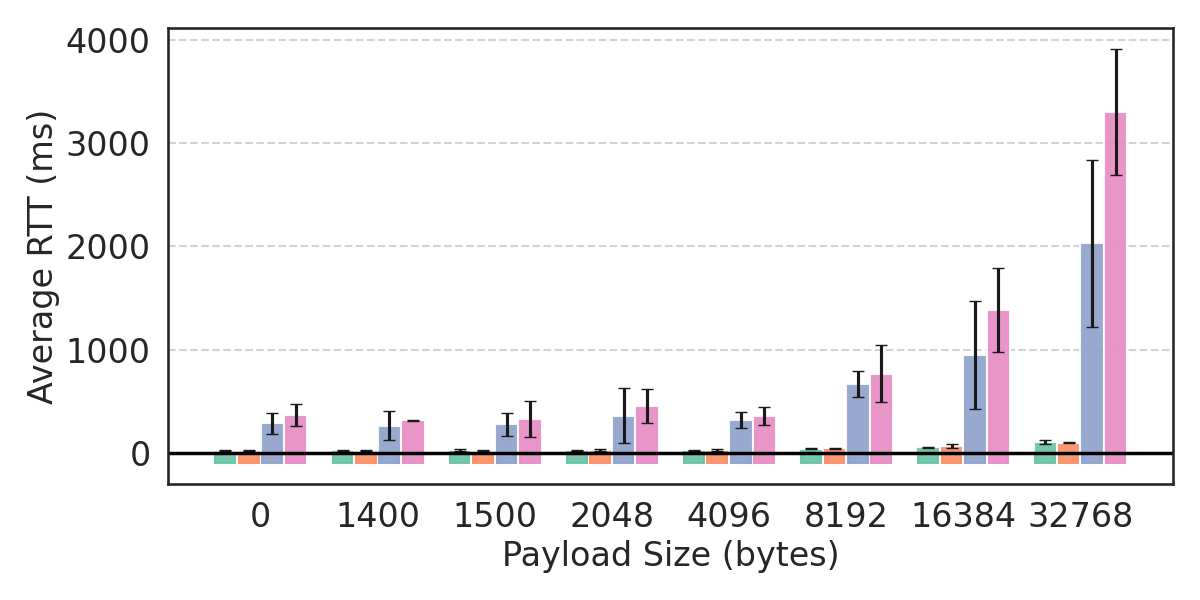}
   \caption{Average RTT at 22.4 Mb/s base-load.}
   \label{fig:wireless_rtt1}
   \end{tcolorbox}
\end{subfigure}
\hfill
\begin{subfigure}[b]{0.325\textwidth}
   \centering
   \begin{tcolorbox}[
       colback=blue!3,
       colframe=blue!30!white,
       arc=2mm,
       boxrule=0.3pt
   ]
   \includegraphics[width=\textwidth]{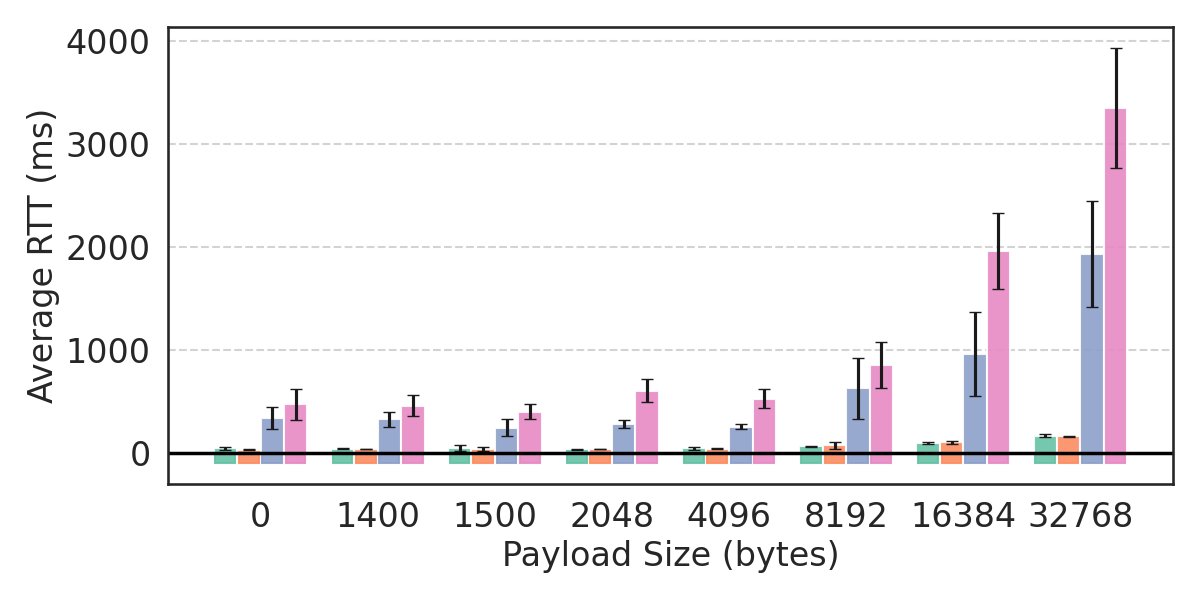}
   \caption{Average RTT at 25.6 Mb/s base-load.}
   \label{fig:wireless_rtt2}
   \end{tcolorbox}
\end{subfigure}
\hfill
\begin{subfigure}[b]{0.325\textwidth}
   \centering
   \begin{tcolorbox}[
       colback=blue!3,
       colframe=blue!30!white,
       arc=2mm,
       boxrule=0.3pt
   ]
   \includegraphics[width=\textwidth]{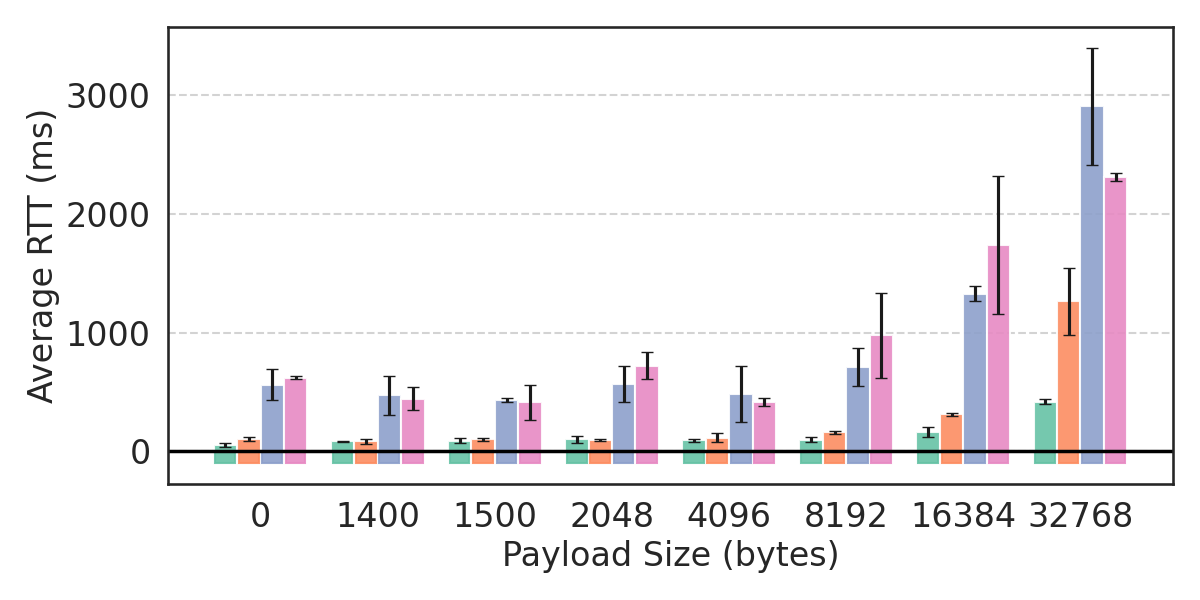}
   \caption{Average RTT at 28.8 Mb/s base-load.}
   \label{fig:wireless_rtt3}
   \end{tcolorbox}
\end{subfigure}
\end{tcolorbox}

\vspace{-3mm}
\begin{center}
   \includegraphics[scale=0.2]{Figures/2025-01-15-ping-legend.png}
\end{center}
\vspace{-3mm}
\caption{Roundtrip ICMP and one-way TCP data in wireless setup}
\label{fig:WirelessEval}
\end{figure*}

In \autoref{fig:WirelessEval} we demonstrate the results of the DoS attack in the wireless setup for TP-Link TL-WA901N. 
As seen, the disruptions were broadly similar to those in the wired setup, with additional vulnerabilities due to the shared nature of transmission queues. In the roundtrip ICMP and one-way TCP data metrics the results were broadly similar to the wired setting, with higher RTTs across the board and even some drops in the most aggressive baseline scenarios. A notable difference is that in the wireless setting, ICMP traffic struggled more than TCP traffic, as it had to traverse the overloaded transmission queue twice.

\begin{figure}[!htb]
\centering
\begin{tcolorbox}[
   colback=gray!5,
   colframe=gray!40!black,
   width=0.95\columnwidth,
   arc=5mm,
   boxrule=0.5pt
]
\begin{subfigure}[b]{\columnwidth}
   \centering
   \begin{tcolorbox}[
       colback=blue!3,
       colframe=blue!30!white,
       arc=2mm,
       boxrule=0.3pt,
       width=0.95\columnwidth
   ]
   \includegraphics[width=\columnwidth]{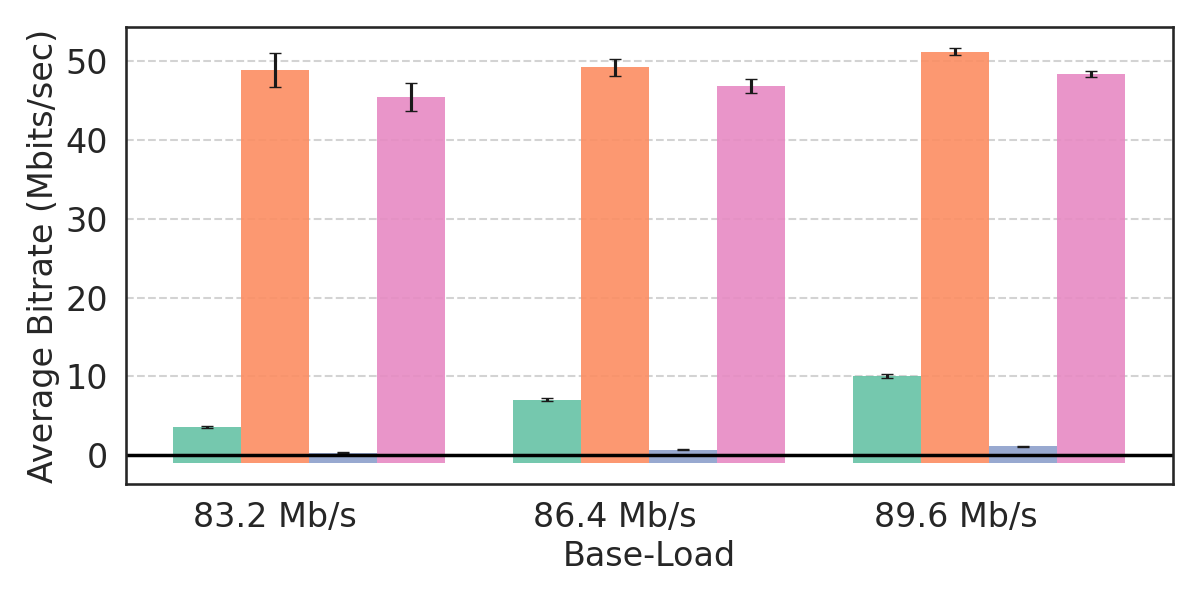}
   \caption{Wired setup}
   \label{fig:iperf_wired}
   \end{tcolorbox}
\end{subfigure}

\vspace{2mm}

\begin{subfigure}[b]{\columnwidth}
   \centering
   \begin{tcolorbox}[
       colback=blue!3,
       colframe=blue!30!white,
       arc=2mm,
       boxrule=0.3pt,
       width=0.95\columnwidth
   ]
   \includegraphics[width=\columnwidth]{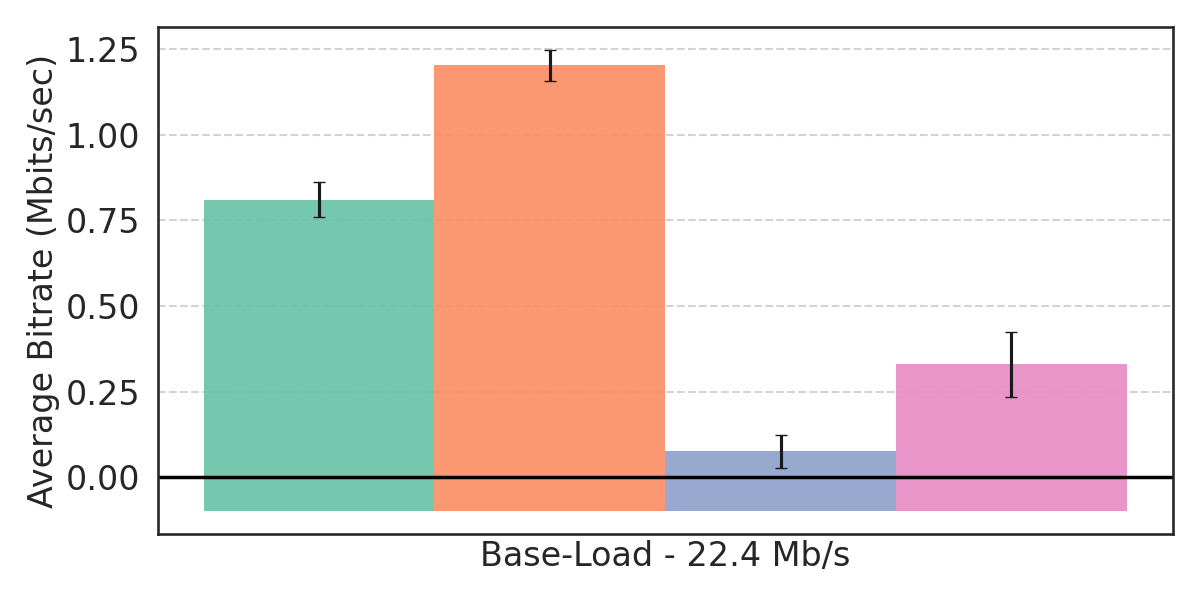}
   \caption{Wireless setup}
   \label{fig:iperf_wireless}
   \end{tcolorbox}
\end{subfigure}
\end{tcolorbox}

\vspace{-3mm}
\begin{center}
\includegraphics[scale=0.2]{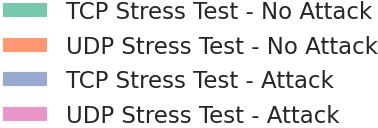}
\end{center}
\vspace{-3mm}
\caption{TCP and UDP stress tests in wired and wireless setups}
\label{fig:iperf}
\end{figure}

Due to the already stressed transmission queue, sending large volumes of traffic became tediously slow before initiating the attack and nearly impossible during it, and so only the TCP and UDP stress test metrics were benchmarked for the lightest base-load setting, in which we see a 90\% reduction in TCP traffic and a 75\% reduction in UDP traffic, as can be seen in \autoref{fig:iperf}. 
The notable reduction in UDP traffic compared to the wired setup is primarily caused by the attack's ability to increase the load on the shared wireless channel, meaning less UDP traffic may be transmitted to the router. 
These findings illustrate that IP cameras with adaptive bitrate can be exploited to cause DoS in wireless networks, particularly in scenarios with limited segmentation and shared resources. As in the wired case, these results occur as soon as the laser is aimed at the camera.

\subsubsection{Impact on Camera-Generated Traffic}

Before commencing the attack, the video transmitted by our camera had an average I-frame size of \textasciitilde 75 KB, and an average P-frame size of \textasciitilde 5 KB. This is indicative of a video directed at a scene rich in detail but with few frame-to-frame differences.

As we direct the laser to dazzle the camera, we note the average I-frame and P-frame sizes are both around \textasciitilde 30 KB. As our camera transmits an I-frame every 120 frames, this corresponds to the previously mentioned \textasciitilde $6 \times$ increase in video size.

From this, we learn that the frames generated during the attack have higher spatial redundancy, but near-zero temporal redundancy, which allows the attack to substantially increase the video's bitrate.

\section{Taxonomy of Multidimensional Cyberattacks}
\label{sec:taxonomy}
\begin{figure*}[tbp]
    \centering
    \includegraphics[scale=0.5]{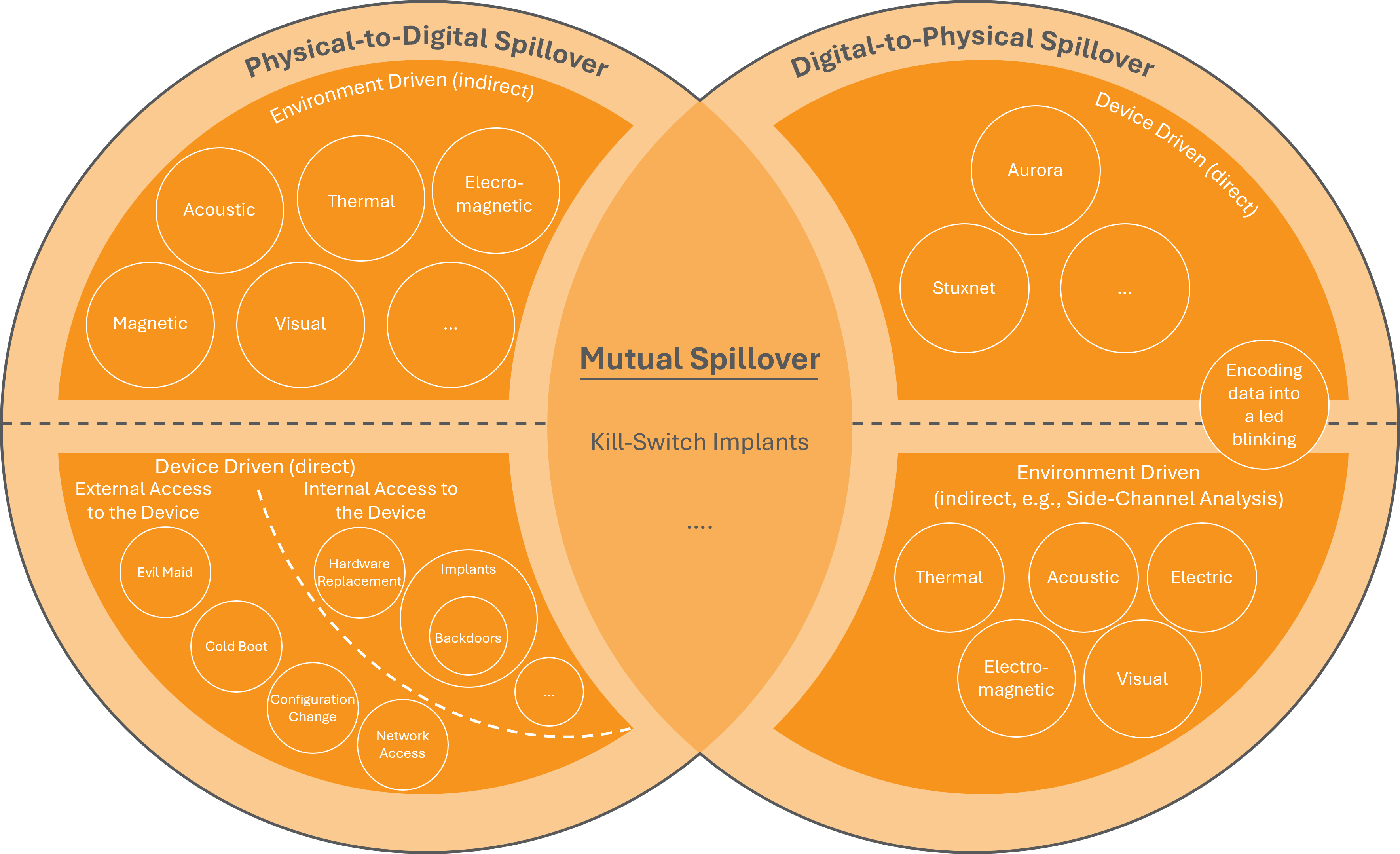}
    \caption{Taxonomy of multidimensional attacks.}
    \label{fig:vienn_taxonomy_attacks}
\end{figure*}

In our view, this work falls under a broader context, of the intersections between the digital and the physical in the context of cybersecurity. 
Intuitively, we can categorize the intersection between the physical and digital in the context of cybersecurity in two main ways: (i) how the digital affects the physical (we call this category "digital-to-physical spillover"), and (ii) how the physical affects the digital (we call this category "physical-to-digital spillover").
This classification is based on the notion that while some events begin in the digital domain and have a physical manifestation, others go in the opposite direction, starting in the physical world while eventually inflicting digital damage.
In the offensive context, we place attacks where the attacker manipulates some aspects of the physical world to inflict damage in the digital domain in the first category. 
An extreme example in this category is a case in which an attacker tries to set a data center on fire, an attack that affects the availability of a web application. 
The second category deals with attacks that emerge from the digital domain intending to inflict physical damage. 
A well-known and much-discussed example of an attack in this category is the Stuxnet malware that physically crippled centrifuges in a nuclear enrichment facility~\cite{langner2013tokill}.
However, this broad categorization is inadequate in light of the diversity of the attacks that have emerged over the years and the increased granularity in the categorization is needed to account for the differences, similarities, and nuances of the attacks. 
We believe that correctly mapping the various attacks in a dedicated taxonomy will help researchers identify gaps and focus their work on a specific sub-domain requiring further attention. 
With this mission in mind, we created a taxonomy outlining multidimensional intersections between the physical and the digital in the context of cybersecurity. 
Since mapping all the various touch points in a single taxonomy may be overwhelming, in this paper, we focus on an offensive view, as presented in \autoref{fig:vienn_taxonomy_attacks}. For completeness of discussion, we briefly describe additional non-offensive perspectives below. 

\subsection{Multidimensional Taxonomy - an Offensive View}

Our offensive taxonomy is comprised of two main parts: (i) \textit{physical-to-digital spillover} where the attacker has physical access to the target, which they leverage to have a digital impact; and (ii) \textit{digital-to-physical spillover} where the attacker has a digital foothold in the target, which they leverage to have a physical impact. 
We elaborate on both in the subsections below.

\subsubsection{Physical-to-Digital Spillover}

This dimension deals with events that begin in the physical domain and spill over to the digital domain \cite{keyboard_acoustic,cpu_fan_tempest,lamphone2020,nassi2020laser,nsa_gunman,laptop_listening2019,photonic_tempest,power_tempest,rambleed2019,kim2014rowhammer,screaming_channels,synesthesia2019,thermal_imaging2018,genkin2014acoustic}. 
It is broken down into two main categories: (i) environment-driven (indirect); and (ii) device-driven (direct). 

\paragraph{Environment-Driven}

Environment-driven attacks exploit external phenomena to fool or overload physical input. 
Because these methods typically do not require disassembly or deep access to the target hardware, they are considered “indirect”. 
Typically, the attacker manipulates environmental factors that are external to the device so that the system’s sensors or operational logic are tricked or overloaded. 
Because these methods typically affect the system indirectly through its environment, attackers do not need to open up or directly handle the device. Instead, they leverage the device’s reliance on ambient signals or physical inputs. 
We split this category into different environmental modalities that the threat actor may try to influence, including the acoustic, thermal, and electromagnetic modalities. 
Examples include spoofing a GPS receiver by broadcasting fabricated coordinates (electromagnetic), or changing the temperature to cause erroneous sensor readings (thermal), etc.

\paragraph{Device-Driven}

In contrast, in device-driven attacks, the attacker exerts direct control over the target device’s hardware rather than controlling the device's environment. 
Unlike environment-driven attacks, which rely on tricking sensors or altering ambient conditions, device-driven compromises allow attackers to directly connect to or modify the device’s functionality at its core. 
We distinguish between two main approaches for device-driven attacks based on how evasive the attacker is when modifying the device:

\begin{itemize} 
    \item \textbf{External Interventions}: 
Here, the attacker manipulates the device without disassembling it or making internal hardware modifications. 
For example, in the classic evil maid attack, an adversary gains physical access to an unattended device such as a laptop and compromises it to facilitate later unauthorized digital access~\cite{boursalian2019bootbandit}.
By booting the device from an external medium (e.g., USB drive) or modifying bootloader settings, the attacker can capture encryption keys or passwords when the device is used.
Typical methods include plugging in malicious peripherals, conducting “cold boot” exploits, or reconfiguring configuration settings. 
In this category, we also include cases where the attacker simply uses the device as an entry into the network, for example, by physically connecting to it and using it as a "jump-box" into the network. Such attacks often exploit accessible ports, removable media, or the device’s default maintenance modes. Although these attacks require physical proximity or hands-on access, they do not necessarily demand deep technical modifications, making them easier to deploy, albeit somewhat less stealthy than fully internal approaches.
\item \textbf{Internal Interventions}:
In contrast, internal intervention attacks require opening up the device and physically altering or adding hardware components. Examples include implanting backdoor modules, swapping out hardware parts (e.g., memory chips), or incorporating malicious circuitry at the board level. Because the attacker directly inserts or replaces hardware, these internal interventions can offer long-term persistence. However, greater technical skill is required to compromise the supply chain or modify the device on-site, increasing both the complexity and the stealth requirements of the operation.

\end{itemize}

\subsubsection{Digital-to-Physical Spillover}

The digital-to-physical spillover dimension deals with events that begin in the digital domain and spill over to the physical domain \cite{guri2022air,guri2014airhopper,weiss2016aurora,colonial2021,guri2017acoustic,guri2016fansmitter,lee2014german,kozak2023industroyer,industroyer2_2022,iranian_steel_2022,guri2021lantenna,guri2017led,guri2021magneto,guri2018mosquito,guri2019odini,oldsmar2021,pipedream2022,guri2019powerhammer,farwell2011stuxnet,di2018triton,guri2016usbee,rippler2018,greenberg2017wannacry,cisco2018olympicdestroyer,greenberg2018notpetya,kaspersky2012shamoon}. 
It includes attacks that begin with software, firmware, or network-based exploits and culminate in tangible, real-world effects such as physical sabotage or exfiltration of data through covert physical channels.
In a sense, this dimension is a mirror view of physical-to-digital spillover, both in terms of the semantic meaning and also the way it is structured. 
Thus, we draw a parallel device-driven (direct) and environment-driven (indirect) distinction:

\begin{itemize} 
    \item \textbf{Device-Driven}:
Device-driven attacks leverage software or network vectors to reprogram, damage, or misuse physical machinery. 
In these types of attacks, the adversary has digital access to a device, and by using this access, they achieve a malicious physical effect. 
Attacks in this category are considered direct because the attacker exerts control over the device and affects the hardware’s physical operations, albeit via digital instructions. 
Accordingly, this dimension examines how the inherent properties of the physical components of computing systems can become direct targets of cyberattacks \cite{NozomiTriton2018, IoTBrick2020}. 
Such attacks can cause physical damage to hardware. 
A notable example in this respect is the notorious Aurora experiment conducted by the U.S. Department of Energy, in which a firmware update caused a power generator to explode by tampering with synchronization frequency logic \cite{greenberg2017_30lines}. 
However, the most widely discussed attack in this category is undoubtedly the Stuxnet malware, which was able to cause physical damage to centrifuges in a nuclear enrichment facility \cite{langner2013tokill}.



    \item \textbf{Environment-Driven}: In contrast, environment-driven attacks in this dimension often fall under the umbrella of side-channel analysis, where attackers measure or induce subtle physical signals of various modalities (e.g., electromagnetic emissions, acoustic cues, or LED flashes) to extract information about the digital processes. 
    In this case, the attacker does not have a digital presence on the target system, but rather exploits the physical byproducts of computations emitted to the environment. In other words, the attacker has some access to the environment of the device, and by capitalizing on the physical properties of the environment, the attacker infers information about the digital processes of the system. For instance, attackers can perform a power analysis to deduce encryption keys, turning an inherent aspect of computing into a potent tool for compromise \cite{Nassi2024}. 
    This dimension was brought to the public's attention by the TEMPEST program \cite{wiki:tempest}. 
Here, the attacker does not directly seize control of the machine’s functions; instead, they rely on inadvertent or engineered leakage between digital processes and the surrounding physical environment.
\end{itemize}

\subsubsection{Intersections}

Some threats can be categorized in both environment-driven and device-driven categories of the same spillover class. Let's take an example attack where the attacker forces an LED of a device to flash in a way that encodes classified documents stored on the device, allowing their exfiltration~\cite{guri2017led}. 
Here, the attacker has to install malware on a device, to cause a physical effect, falling under the device-driven category of the digital to physical spillover. Next, the attacker uses the fact that they have a line-of-sight to this LED to obtain the encoded information, falling under the environment-driven category. 
This attack blurs the boundaries between the categories in the same spillover class.

Moreover, some threats defy easy categorization as strictly physical-to-digital or digital-to-physical, particularly those that can be triggered and exploited in both domains. 
We refer to these as mutual spillover techniques. 
Mutual spillover attacks include cases where the direction of the spillover between physical and digital domains is blurred. 
A classic example is a kill-switch implant, which might be embedded in hardware at manufacture time and can be activated either by a physical trigger (e.g., applying a magnet or shorting specific pins) or a digital command (e.g., receiving a coded message over the network). 

Similarly, an attacker who physically controls a water monitoring sensor could repeatedly dip it into two different buckets of water with different content profiles, causing the sensor to perpetually sample and transmit data.

\subsection{Additional Spillovers Between the Physical and the Digital in the Cybersecurity Context}

We note that looking only at offensive and spillovers between the physical and the digital does not tell the entire story. For example, future research could address the defensive perspective, highlighting defensive strategies that leverage the physical properties of computing systems. This perspective involves detecting anomalies in physical signals (such as power consumption, electromagnetic emissions, and packet delay) to identify anomalies \cite{reed2012enhancing, chen2017powerful, clark2013wattsupdoc,ding2020deeppower,lightbody2023attacks,xu2019addp,li2019using,zhang2023trustguard}. For example, defenders can establish baseline profiles of power usage for various legitimate computational processes and monitor for deviations indicative of abnormal activity, thereby utilizing the physical byproducts of computing as indicators of breaches ~\cite{munny2021power, clark2013wattsupdoc, lightbody2023attacks, xu2019addp,li2019using, zhang2023trustguard}.

Another, less explored spillover between the physical and the digital in the cybersecurity context deals with the measurement and optimization of energy consumption by cybersecurity solutions themselves \cite{GreenSec2024}. This aspect examines how the very tools and processes designed to protect computing systems can also place a significant demand on physical resources, such as energy consumption ~\cite{annual_co2_emmisions_worldwide, ict_and_cybersecurity_energy_consumption, powerslave_measure_6_antiviruses, energy_efficient_security_survey, rootkits}. Conversely, the scarcity of physical resources may also influence the design and operational patterns of cybersecurity solutions running on top of a resource-constrained device, like an Internet of Things (IoT) device or a satellite system \cite{salim2024cybersecurity}. This aspect of cybersecurity calls for a balanced approach, weighing the effectiveness of security controls against their energy consumption to ensure that protective measures align with broader goals of energy efficiency and sustainability.

\subsection{Mapping Our Proposed Attack on the Taxonomy}

While crude physical attacks, such as destroying a data center, can certainly disrupt digital systems, our taxonomy enables a more nuanced approach. Traditional cyber-physical targets like data centers are typically hardened against physical attacks through physical security measures such as guards, access controls, and surveillance systems. However, the growing deployment of IoT devices and sensors has created a new and more accessible attack surface for adversaries. These devices are often installed in public or semi-public spaces, where physical access cannot be completely restricted. For instance, security cameras are commonly mounted on external walls or poles to maximize their field of view, making them physically accessible to attackers who can manipulate their surroundings without needing direct contact or causing obvious damage. Capitalizing on this insight, we demonstrate how attackers can exploit the inherent physical-to-digital conversion properties of network-connected devices, such as IP cameras, to conduct denial-of-service attacks.
While attacks like Stuxnet~\cite{farwell2011stuxnet} and Aurora~\cite{weiss2016aurora} exploited the digital domain to cause physical damage, we demonstrate how physical properties can be weaponized to inflict damage in the digital realm. In taxonomy terms, we explore a new attack in which an attacker can manipulate physical-world artifacts to inflict damage in the digital domain. Namely, since the attacker changes the scene that the camera sees, this attack falls under the visual modality of the environment-driven category of the physical-to-digital spillover attack dimension.

\section{Discussion}
\label{sec:discussion}


This article explored the new laser-generated DoS attack, which caused video surveillance cameras to generate excess traffic and overload their local network. 
Combining the attack approach described in this article with carefully crafted rapid scene manipulations described by ~\cite{nassi2018gamedronesdetecting}, future research can explore stealthy versions of laser-based DoS without camera dazzling.

The effects of the attack could be mitigated using controls at the network level and at the host level. 
At the network level, physical network segmentation provides the strongest protection by isolating resources dedicated to handling critical traffic from resources used to deliver the camera traffic. 
That said, a lack of network segmentation continues to plague the world in general and cyber-physical installations in particular. According to a recent survey, 77\% of service engagements of a leading ICS security consultancy and incident response firm involved issues with network segmentation \cite{dragos2021cybersecurity}. A more disturbing finding we came across is that in offshore platforms, security cameras often share network equipment with industrial control elements \cite{siemens2019reliable}.
In cases where segmentation is done virtually rather than physically (e.g., via VLANs), IP cameras can still be a hazard as they can overload the shared physical resources of the network, regardless of virtual segmentation. 

Several network equipment vendors offer a security control of port rate limitation where ingress ports can be configured to drop incoming traffic above a certain threshold. 
This feature could be useful to prevent the cameras from generating excessive network traffic.
At the host level, several camera vendors allow configuring their devices in a way that can contain excessive traffic; configuring cameras to transmit at a constant bitrate can prevent potential traffic spikes.

Finally, physical security controls preventing attackers from approaching the camera at a line of sight could also be considered. 
However, in most cases, the cameras themselves are deployed as a physical control to detect trespassers.

\section{Conclusions}

This paper documents how attackers can manipulate IP camera behavior to disrupt networks. 
The attack works by using a laser diode to dazzle an IP camera, drastically reducing the effectiveness of video compression algorithms. 
Through this physical manipulation, the attack is causing IP cameras to generate high-bandwidth traffic.
This attack represents a broader class of attacks where the threat actor can modify the physical environment in a way that triggers variable bitrate devices to generate excessive network traffic.
The practical implications of this inherent property require attention, particularly in industrial settings where network disruptions can stop operations. 
Network security best practices, such as segmentation and rate limiting, offer strong protection against this attack.


\section*{Ethics Considerations} 

Through this research, we aim to raise awareness about defensive gaps in cyber-physical systems and promote the implementation of protective measures. Our research improves cybersecurity by mapping the intersections between physical and digital attack vectors to shed light and advance defensive research in an underexplored area. 

Only one trial was performed outside the lab to measure the effective attack distance.
Our paper deliberately omits these details alongside some other attack implementation details that could contribute to malicious use, while providing enough technical information for organizations to understand and defend against these threats. 
Since laser radiation may endanger the beholder, we did not perform full experiments outside the lab out of caution and care for the health and safety of the researchers and passersby.

The defenses we propose are practical and cost-effective; in addition, they can typically be implemented with existing equipment and, therefore, do not require investment in new equipment. Most organizations can implement our recommended protective controls using their existing network infrastructure through configuration changes and security policy updates. These defensive measures, including network segmentation, traffic management, and monitoring systems, also improve robustness against other attacks, making networks more resilient overall.

\section*{Open Science} 

This research did not produce any datasets, scripts, binaries, nor any significant source code. All hardware and software utilities used during this research are common open-source and off-the-shelf items, described in this paper.

\bibliographystyle{unsrt}
\bibliography{sample}

\end{document}